\documentclass[letterpaper,11pt]{article}
 \pdfoutput=1
\usepackage{jheppub}

\usepackage{amssymb,amsmath,amssymb,graphicx,epsfig,slashed,hyperref,doi}

\newcommand{\be}{\begin{equation}}
\newcommand{\ee}{\end{equation}}
\newcommand{\bea}{\begin{eqnarray}}
\newcommand{\eea}{\end{eqnarray}}
\newcommand{\beq}{\begin{equation}}
\newcommand{\eeq}{\end{equation}}

\def\circa#1{\,\raise.3ex\hbox{$#1$\kern-.75em\lower1ex\hbox{$\sim$}}\,}

\begin{document}

\title{Vector SIMP dark matter}

\author[1]{Soo-Min Choi,}
\author[2,3]{Yonit Hochberg,}
\author[2,3]{Eric Kuflik,}
\author[1]{Hyun Min Lee,}
\author[4]{Yann Mambrini,}
\author[5,6,7,8]{Hitoshi Murayama,}
\author[4]{Mathias Pierre}

\affiliation[1]{Department of Physics, Chung-Ang University, Seoul 06974, Korea}
\affiliation[2]{Department of Physics, LEPP, Cornell University, Ithaca NY 14853, USA}
\affiliation[3]{Racah Institute of Physics, Hebrew University of Jerusalem, Jerusalem 91904, Israel}
\affiliation[4]{Laboratoire de Physique Th\'eorique (UMR8627), CNRS, Univ. Paris-Sud,   Universit\'e Paris-Saclay, 91405 Orsay, France}
\affiliation[5]{Ernest Orlando Lawrence Berkeley National Laboratory, University of California, Berkeley, CA, 94720, USA}
\affiliation[6]{Department of Physics, University of California, Berkeley, CA 94720, USA}
\affiliation[7]{Kavli Institute for the Physics and Mathematics of the Universe (WPI), University of Tokyo
Institutes for Advanced Study, University of Tokyo, Kashiwa 277-8583, Japan}
\affiliation[8]{Center for Japanese Studies, University of California, Berkeley, CA 94720, USA}

\preprint{LPT-Orsay-17-10, IPMU17-0094}

\abstract{Strongly Interacting Massive Particles (SIMPs) have recently been proposed as light thermal dark matter relics. Here we consider an explicit realization of the SIMP mechanism in the form of vector SIMPs arising from an $SU(2)_X$ hidden gauge theory, where the accidental custodial symmetry protects the stability of the dark matter.
We propose several ways of equilibrating the dark and visible sectors in this setup.  In particular, we show that a light dark Higgs portal can maintain thermal equilibrium between the two sectors, as can a massive dark vector portal with its generalized Chern-Simons couplings to the vector SIMPs, all while remaining consistent with experimental constraints.}

\maketitle

\section{Introduction}\label{sec:intro}

There is overwhelming evidence for the existence of dark matter (DM) in the Universe~\cite{Ade:2015xua}. One of the most compelling particle physics candidates for dark matter is the Weakly Interacting Massive Particles (WIMPs). However,  the absence of experimental signals in direct~\cite{Arcadi:2017kky,Aprile:2016swn,Akerib:2016vxi,Tan:2016zwf} and indirect~\cite{Ackermann:2015tah,Abdallah:2016ygi} detection
experiments
for WIMPs, has led researchers to focus attention in recent years on sub-GeV dark matter. Thermal production of such light dark matter is possible if, for instance, standard $2\to2$ annihilations proceed with small couplings~\cite{Feng:2008ya} or if new annihilation mechanisms are present, such as $3\to2$ annihilations~\cite{Carlson:1992fn,Hochberg:2014dra,Hochberg:2014kqa} or forbidden $2\to2$ channels~\cite{Griest:1990kh,DAgnolo:2015ujb}.

The thermal production of Strongly Interacting Massive Particles (SIMPs)~\cite{Hochberg:2014dra} is based on freezeout of $3\rightarrow 2$ self-annihilation of dark matter, with  coupling between SIMPs and light Standard Model (SM) particles, which maintain kinetic equilibrium between the two sectors until  freeze-out occurs.  Various realizations of SIMP dark matter have been proposed in the literature, which often contain (pseudo)scalar dark matter particles with dark abelian or non-abelian gauge symmetries~\cite{Hochberg:2014dra,Lee:2015gsa,Choi:2015bya,Hochberg:2015vrg,Choi:2016hid,Choi:2016tkj,Choi:2017mkk}.
Massive dark vector bosons can also be SIMP candidates when stemming from non-abelian dark gauge bosons~\cite{Hambye:2009fg,Hambye:2008bq,Karam:2015jta,Karam:2016rsz,Bernal:2015ova,Heikinheimo:2017ofk}, as can be dark fermions or scalars when accompanied with a light dark photon or another scalar ~\cite{Cline:2017tka,Dey:2016qgf}.
Vector SIMP models are particularly predictive since the cubic and quartic self-interactions of dark matter are determined by a single gauge coupling.
If the non-abelian dark gauge symmetry is spontaneously broken by the Higgs mechanism, the resulting massive dark Higgs can equilibrate the vector SIMPs and the SM via a Higgs portal coupling~\cite{Hambye:2009fg,Hambye:2008bq,Karam:2015jta,Bernal:2015ova,Kamada:2016ois}. The spin information of the dark matter could be then be inferred from the invisible Higgs decay, as is the case for the WIMP~\cite{Lebedev:2011iq}.

In this paper, we consider vector SIMP dark matter in an $SU(2)_X$ dark gauge theory, where the three massive (degenerate) $SU(2)_X$ gauge bosons play the role of vector SIMPs.
Equilibration between the dark and visible sectors can be achieved by elastic scattering between the dark matter and the $SU(2)_X$ dark Higgs, provided that the latter is light enough to be thermalized with the SM via the Higgs portal until freeze-out occurs. As we will see, the dark Higgs can successfully thermalize the two sectors only when it is close in mass to the dark matter, in which case additional forbidden $2\to2$ annihilations are important as well.
Alternatively, the dark $U(1)_{Z'}$ photon can thermalize the dark and visible sectors via its kinetic mixing with the SM hypercharge alongside its coupling to the DM, which proceed through generalized Chern-Simons (CS) terms~\cite{Anastasopoulos:2006cz,Antoniadis:2009ze,Mambrini:2009ad,Dudas:2009uq,Kim:2015vba,Arcadi:2017jqd}. In both cases of the Higgs and vector portals, we find parameter space consistent with all existing constraints. 
Our results indicate that the framework can be probed via Higgs/$Z$-boson invisible decays as well as dark Higgs/dark photon searches in current and future collider and beam dump experiments.

This paper is organized as follows.
We begin by describing the $SU(2)_X$ dark gauge theory model in Section~\ref{sec:model}, including the relevant Higgs and gauge-mixing vector portals to the SM.
Section~\ref{sec:relic} discusses the $3\rightarrow 2$ annihilation processes setting the DM abundance, the self-scattering cross sections, and the effects of forbidden channels on the relic density.
Methods for achieving kinetic equilibrium between the dark and visible sectors via Higgs mixing and/or gauge mixing are addressed in Section~\ref{sec:portal}. We conclude in Section~\ref{sec:conc}.

\section{The model}\label{sec:model}

Here we present the framework for vector SIMPs: We start with the dark gauge theory, and then describe the Higgs interactions as well as kinetic gauge mixing and couplings between the dark photon and the dark matter.

\subsection{The dark sector}\label{ssec:darksector}

We consider as a model for non-abelian SIMP dark matter an $SU(2)_X$ gauge theory in the dark sector, broken completely due to the VEVs of a dark Higgs doublet $\Phi$.  The massive gauge bosons of $SU(2)_X$, denoted by $X^i_\mu$ $(i=1,2,3)$, are degenerate and stable due to a dark custodial isospin symmetry, and are a dark matter candidate~\cite{Hambye:2009fg,Hambye:2008bq,Bernal:2015ova}. The accidental custodial symmetry persists in the presence of the Higgs portal and $Z'$ portal with the generalized Chern-Simons term which we discuss later, maintaining the stability of the dark matter.

The Lagrangian for the dark sector is given by
\be
{\cal L}= -\frac{1}{4} {\vec X}_{\mu\nu}\cdot {\vec X}^{\mu\nu}+ \mathcal{L}_{\rm scalar}\,,
\ee
where the field strength tensors are ${\vec X}_{\mu\nu}=\partial_\mu {\vec X}_\nu-\partial_\nu {\vec X}_\mu+g_X ({\vec X}_\mu\times {\vec X}_\nu) $. The scalar potential is given by
\bea\label{eq:Vhiggs}
\mathcal{L}_{\rm scalar}&=&   |D_\mu \Phi|^2 +m^2_\Phi |\Phi|^2 -\lambda_\Phi |\Phi|^4\,,
\eea
with the covariant derivatives for the dark Higgs doublet is $D_\mu \Phi=(\partial_\mu-\frac{1}{2} i g_X {\vec\tau}\cdot {\vec X}_\mu)\Phi$.

After expanding the dark Higgs fields around the VEV as $\Phi=\frac{1}{\sqrt{2}}(0,v_X+ \phi)^T$ in unitary gauge, one obtains gauge boson mass of $m_X=\frac{1}{2} g_X v_X$. The self-interactions of the vector dark matter and its interactions with the dark Higgs $\phi$ are given by
\bea
{\cal L} &\supset&-\frac{1}{2}g_X (\partial_\mu {\vec X}_\nu -\partial_\nu {\vec X}_\mu)\cdot ({\vec X}^\mu\times {\vec X}^\nu) -\frac{1}{4} g^2_X ({\vec X}_\mu\cdot {\vec X}^\mu)^2 \nonumber \\
&&+\frac{1}{4}g^2_X({\vec X}_\nu\cdot {\vec X}^\mu)({\vec X}_\mu\cdot {\vec X}^\nu)+\frac{1}{2} m^2_X {\vec X}_\mu \cdot {\vec X}^\mu \left(\frac{2\phi}{v_X}+\frac{\phi^2}{v^2_X} \right)\,. \label{darkHiggs}
\eea

The non-abelian interactions among the vector bosons $X$ allow for $3\rightarrow 2$ annihilations as SIMPs.  This idea is actually much more general than we discussed above.  This symmetry breaking can also be considered as dynamical, as a result of chiral symmetry breaking $SU(2)_L \times SU(2)_R \rightarrow SU(2)_V$ in an $SU(N_c)$ gauge theory.  This corresponds to the limit where $m_\phi \rightarrow \infty$ at low energies, while resonances can play an important role at higher energies.  In this case, the coupling $g_X$ is still considered perturbative.

Alternatively, we can consider the theory with a Higgs doublet in the strongly coupled regime $g_X \gg 1$.  As pointed out by 't Hooft~\cite{tHooft:1979yoe}, an $SU(2)$ gauge theory with a doublet scalar does not have an order parameter to distinguish the broken and confining phases, and hence the two phases are continuously connected, akin to liquid and gas phases of water at high pressures.  In the strong coupling case, the vector SIMP is described by the interpolating field $\Phi^\dagger i\mathop{D}\limits^{\leftrightarrow}{}_\mu \Phi$, while the dark Higgs by $\Phi^\dagger \Phi$.  Given enough parameters in the model $(g_X, m_\Phi^2, \lambda_\Phi)$, one can most likely have the dark Higgs heavier than the vector SIMP as required (see below); such a discussion requires numerical simulations and is beyond the scope of this paper.

\subsection{Higgs portal}\label{ssec:higgs}
The dark Higgs provides a portal between the dark sector and the visible sector, since the dark and SM scalars may interact at the renormalizable level,
\beq
{\cal L}_{\rm higgs} =  \lambda_{\Phi H}|\Phi|^2|H|^2+\lambda_{SH}|S|^2 |H|^2+ \lambda_{\Phi S}|\Phi|^2|S|^2\,.
\eeq
Here, a complex scalar field $S$ is introduced for giving mass to $Z’$ gauge boson by Higgs mechanism in the later discussion on Z’ portal in Sec. 2.3. Since $Z’$ is assumed to be heavier than dark matter in our model, we assumed that the radial mode of $S$ has no significant mixing with the dark Higgs $\phi$ and the SM Higgs.

The SM and dark Higgs bosons are then mixed by
\be\label{eq:hmix}
\left( \begin{array}{c} h_1 \\ h_2 \end{array} \right)=  \left( \begin{array}{cc} \cos\theta & -\sin\theta \\  \sin\theta & \cos\theta \end{array} \right) \left( \begin{array}{c} \phi \\ h \end{array} \right)\,,
\ee
where $h_1,h_2$ are mass eigenstates of mass
\be
m^2_{h_1,h_2}=\lambda_\Phi v^{ 2}_X+\lambda_H v^2 \mp \sqrt{(\lambda_\Phi v^{ 2}_X-\lambda_H v^2)^2+ \lambda^2_{\Phi H}v^{2}_X v^2 }\,,
\ee
and the mixing angle is given by
\be
\tan 2\theta = \frac{\lambda_{\Phi H} v_X v}{\lambda_H v^2-\lambda_\phi v^2_X}.
\ee
Here, we assume that the additional Higgs field $s$ for $U(1)_{Z'}$ is heavy enough so that its mixing effects with the above Higgs fields is negligible.
The Higgs mixing yields interactions between the vector DM and the SM particles,
\bea
{\cal L}&\supset&\frac{m^2_X}{v_X}\, {\vec X}_\mu \cdot {\vec X}^\mu (\cos\theta\, h_1+\sin\theta\, h_2)+\frac{m^2_X}{2v^2_X}\, {\vec X}_\mu \cdot {\vec X}^\mu (\cos\theta\, h_1+\sin\theta\, h_2)^2 \nonumber \\
&&-\frac{m_f}{v}\, {\bar f} f (-\sin\theta\, h_1+\cos\theta\, h_2)\,,
\eea
enabling communication between the two sectors.

In the presence of such Higgs-portal couplings, the SM Higgs can decay invisibly into a pair of dark gauge bosons or dark higgses, with decay rates
\bea\label{eq:h2decay}
\Gamma(h_2\rightarrow X X)&=& \frac{3\sin^2\theta m^3_{h_2}}{32 \pi v^2_X}\,\bigg(1-\frac{4m^2_X}{m^2_{h_2}}+\frac{12 m^4_X}{m^4_{h_2}} \bigg) \sqrt{1-\frac{4m^2_X}{m^2_{h_2}}}\,,\nonumber \\
\Gamma(h_2\rightarrow h_1 h_1)&=& \frac{\lambda^2_{\Phi H} v^2}{32\pi m_{h_2}}\sqrt{1-\frac{4m^2_{h_1}}{m^2_{h_2}}}\,.
\eea
The visible decays of the SM Higgs are scaled down universally by $\cos^2\theta$ due to the Higgs mixing.  As we will see in Section~\ref{ssec:higgsportal}, the bound on invisible Higgs decays places a strong constraint on the allowed mixing, and hence on the possibility that the Higgs portal maintains kinetic equilibrium between the two sectors.

\subsection{Vector portal}\label{ssec:gauge}

In addition to the Higgs portal, we can gauge a $U(1)_{Z'}$ symmetry acting on the complex scalar $S$, with the covariant derivative $D_\mu S=(\partial_\mu -ig_{Z'}Z'_\mu )S$.  The $U(1)_{Z'}$ massive gauge boson $Z'$ can connect the dark and visible sectors, in the presence of gauge kinetic mixing with the SM hypercharge as well as DM-$Z'$ interactions:
\beq
\mathcal{L}_{\rm vector} = -\frac{1}{2} \,\sin\xi\, Z'_{\mu\nu} B^{\mu\nu}+{\cal L}_{XXZ^\prime}\,.
 \eeq
Here ${\cal L}_{XXZ^\prime}$ generates a 3-pt interaction between $XXZ^\prime$; it may be generated by a non-abelian Chern-Simons (CS) term, as will be discussed below.

The kinetic and mass terms for the $Z$ and $Z'$ gauge bosons~\cite{Choi:2015bya} is diagonalized
\be
\left(\begin{array}{c} B_\mu \\ W^3_\mu \\ Z'_\mu \end{array}\right)=\left( \begin{array}{ccc} c_W & -s_W c_\zeta +t_\xi s_\zeta  & -s_W s_\zeta-t_\xi c_\zeta  \\   s_W & c_W c_\zeta & c_W s_\zeta   \\  0 & -s_\zeta/c_\xi & c_\zeta/ c_\xi   \end{array} \right)  \left(\begin{array}{c} A_\mu \\ Z_{1\mu} \\ Z_{2\mu} \end{array}\right)
\ee
where $(B_\mu, W^3_\mu,Z'_\mu)$ are hypercharge, neutral-weak and dark gauge fields, $(A_\mu, Z_{1\mu},Z_{2\mu})$ are mass eigenstates, and $s_W\equiv \sin\theta_W, c_W\equiv \cos\theta_W$, etc.  Here, $Z_{1}$ is $Z$-boson-like and $Z_{2}$ is $Z'$-boson-like, with masses
\bea
m^2_{1,2}= \frac{1}{2}\left[m^2_Z (1+s^2_W t^2_\xi)+m^2_{Z'}/c^2_\xi\pm \sqrt{(m^2_Z(1+s^2_W t^2_\xi)+m^2_{Z'}/c^2_\xi)^2- 4m^2_Z m^2_{Z'} /c^2_\xi} \,\right]\,,
\eea
where the mixing angle is
\be
\tan 2\zeta =\frac{m^2_Z s_W \sin 2\xi}{m^2_{Z'}-m^2_Z(c^2_\xi-s^2_W s^2_\xi )}\,.
\ee
The electromagnetic and neutral-current interactions are then
\bea
{\cal L}_{\rm EM/NC}&=& e A_\mu J^\mu_{\rm EM} + Z_{1\mu} \bigg[ e \varepsilon J^\mu_{\rm EM}+\frac{e}{2s_Wc_W} (c_\zeta-t_W \varepsilon/t_\zeta  )J^\mu_Z -g_{Z'} \frac{s_\zeta}{c_\xi} J^\mu_{Z'} \bigg] \nonumber\\
&&+Z_{2\mu}  \bigg[- e \varepsilon J^\mu_{\rm EM}+\frac{e}{2s_Wc_W} (s_\zeta+t_W\varepsilon)J^\mu_Z +g_{Z'} \frac{c_\zeta}{c_\xi} J^\mu_{Z'} \bigg]\,,
\eea
 where $\varepsilon\equiv c_W t_\xi c_\zeta\simeq c_W \xi$ for $|\xi|\ll 1$, and $J^\mu_{\rm EM}$, $J^\mu_Z$ and $J^\mu_{Z'}$ are electromagnetic, neutral and dark currents, respectively.
 For $m_{Z'}\ll m_Z$, one has $\zeta\simeq -s_W \xi=-t_W \varepsilon$, so the neutral current interaction of the dark photon is negligible due to $s_\zeta+t_W\varepsilon \simeq \zeta+ s_W\xi\simeq 0$.

There are no direct couplings between the SM and the non-abelian vector dark matter at the renormalizable level, because of the non-abelian gauge symmetry. Likewise, there are no direct renormalizable interactions between the $Z^\prime$ and the $X$-boson, since the dark Higgs are not charged under both symmetries (in other words, the dark Weinberg angle vanishes).

 If heavy fermions charged under both $SU(2)_X$ and $U(1)_{Z'}$ are present in the theory, they may generate low-energy effective $XXZ^\prime$ interactions via triangle diagrams. From the effective theory point of view these may manifest as generalized non-abelian Chern-Simons terms~\cite{Dudas:2009uq,Kim:2015vba},
\be
{\cal L}_{\rm CS,EFT} \supset c_1 \epsilon^{\mu\nu\rho\sigma} Z'_\mu {\vec X}_\nu\cdot (\partial_\rho {\vec X}_\sigma - \partial_\sigma {\vec X}_\rho)  \label{CS1}\,.
\ee
Although the coefficient $c_1$ is dimensionless, these are non-renormalizable operators and arise from gauge dimension-8 operators, known as D'Hoker-Farhi terms~\cite{DHoker:1984izu},
\bea
{\cal L}_{\rm CS}&\supset & \frac{i}{M^4} S^\dagger D^\mu S  (D^\nu \Phi)^\dagger {\tilde X}_{\mu\nu}\Phi+{\rm c.c.}
\eea
Likewise, an effective 3-pt interaction can be generated by the gauge invariant dimension-8 operator of the form
\bea
{\cal L}_{D8}= \frac{1}{M^4} \, \epsilon^{\mu\nu\rho\sigma} (\Phi^\dagger X_{\mu\nu} D_\lambda \Phi) \partial^\lambda Z'_{\rho\sigma}\,.
\eea

In this work we will consider the phenomenology of the effective operator Eq.~(\ref{CS1}); Appendix~\ref{app:CS} contains a concrete example of generating the effective Chern Simons term.

We remark on the invisible decays of $Z$ and $Z'$ bosons in our setup.
The $Z$ boson can decay invisibly into a pair of vector dark matter particles through the generalized CS terms in the presence of a gauge kinetic mixing between $Z'$ and $Z$ bosons. But, if $N_f$ heavy fermions $f$ running in triangle diagrams are relatively light  for a sizable CS term (but heavy enough not to affect our discussion on vector SIMPs in the later sections) as discussed in Appendix~\ref{app:CS}, the $Z$-boson preferentially decays directly into a pair of heavy fermions at tree level. Then, the corresponding $Z$-boson invisible decay width is given by
\bea
\Gamma(Z_1\rightarrow f{\bar f})=\frac{N_f \alpha_{Z'} \varepsilon^2m_{Z}}{3 c^2_W}\left(1+\frac{2m^2_f}{m^2_{Z'}}\right)  \left(1-\frac{4m^2_f}{m^2_Z} \right)^{1/2}  \label{Z-inv}
\eea
with $\alpha_{Z'}\equiv g^2_{Z'}/(4\pi)$.
On the other hand, if the heavy fermions are heavier than $m_{Z'}/2$, the  $Z'$ boson decays into a pair of vector dark matter particles via the CS term, with the width
\bea
\Gamma(Z_2\rightarrow XX)= \frac{c^2_1 m^3_{Z'}}{8\pi m^2_X} \left(1-\frac{4m^2_X}{m^2_{Z'}} \right)^{5/2}. \label{Zp-inv}
\eea

\section{Vector SIMP dark matter}\label{sec:relic}

Having established the interactions of the framework, we now address the cross section for the dark matter relic abundance and self-scatterings.
We first determine the relic density of dark matter from $3\to2$ processes in Section~\ref{ssec:SIMP}, and discuss the role of additional forbidden annihilation channels in Section~\ref{ssec:forb}.

\subsection{SIMP channels}\label{ssec:SIMP}

Here we compute the relic density assuming the $3\to2 $ annihilation processes are the dominant number-changing processes. In the presence of an isospin symmetry for the vector dark matter, all components of dark matter have the same mass, and can be treated as identical particles.
Assuming the dark matter remains in kinetic equilibrium with the SM until the time of freeze-out,
the Boltzmann equation for the vector dark matter is given by~\cite{Hochberg:2014dra}
\bea
\frac{d n_{\rm DM}}{dt}+ 3 H n_{\rm DM}&=& -\Big(\langle\sigma v^2\rangle_{3\rightarrow 2}-\langle\sigma v^2\rangle^h_{3\rightarrow 2}\Big) \Big(n^3_{\rm DM}- n^2_{\rm DM}n^{\rm eq}_{\rm DM}\Big) \nonumber \\
&&-\langle\sigma v^2\rangle^h_{3\rightarrow 2}\Big(n^3_{\rm DM}- n_{\rm DM}(n^{\rm eq}_{\rm DM})^2 \Big)\,.
\eea
Here, the thermally averaged $3\rightarrow 2$ annihilation cross-section (away from a resonance) is given by
\bea
\langle\sigma v^2\rangle_{3\rightarrow 2}&=&\frac{25\sqrt{5}g_X^6}{23887872\pi m_X^5}\frac{1}{(m_{h_1}^2-4m_X^2)^2(m_{h_1}^2+m_X^2)^2} \bigg(14681m_{h_1}^8-87520m_{h_1}^6m_X^2 \nonumber \\
&&+21004m_{h_1}^4m_X^4+327580m_{h_1}^2m_X^6 +290775m_X^8\bigg) +\langle\sigma v^2\rangle^h_{3\rightarrow 2}
\label{3to2}
\eea
with
\bea
\langle\sigma v^2\rangle^h_{3\rightarrow 2}=  \frac{\sqrt{5}g_X^6m^{16}_{h_1}}{80621568\pi m_X^{10}} \frac{(1-m^2_{h_1}/(16 m^2_X) )^{1/2} }{(m_{h_1}^2-4m_X^2)^{7/2}(m_{h_1}^2+2m_X^2)^2}\bigg(C_1+\frac{2C_2 m^4_{h_1}}{(m^2_{h_1}-7m^2_X)^2} \bigg)  
\eea
where $C_1$ and $C_2$ are dimensionless quantities given in Eqs.~(\ref{C1}) and (\ref{C2}), respectively.  We note that the first term in $\langle\sigma v^2\rangle_{3\rightarrow 2}$ stems from $XXX\rightarrow XX$ channels and $\langle\sigma v^2\rangle^h_{3\rightarrow 2}$ due to $XXX\rightarrow X h_1$ channels contributes only for $m_{h_1}<2m_X$, becoming dominant near the resonance at $m_{h_1}=2m_X$. On the other hand, $XXX\rightarrow h_1 h_1$ channels are $p$-wave suppressed so they are not included here.  
Additional terms that give an approximate resonance when $m_{h_1}= 3 m_X$ are present, but as they are $p$-wave suppressed they are always subdominant and hence can be neglected. Further details of the $3\rightarrow 2$ cross section and discussion of the Boltzmann equation can be found in Appendix~\ref{app:xsec}.

In the instantaneous freeze-out approximation, the relic abundance for $3\rightarrow 2$ annihilation is found to be
\bea
\Omega_{\rm DM}  \simeq \frac{m_X s_0/\rho_c}{s({m_X})^2 / H(m_X)} \frac{x_f^2}{\sqrt{\langle\sigma v^2\rangle_{3\rightarrow 2}}}\,,
\eea
where ${s_0/\rho_c \simeq 6\cdot 10^8 /\rm GeV}$ is the ratio of the entropy density today to the critical density, $s(m_X)$ is the entropy density at $T= m_X$, and $H(m_X)$ is the Hubble rate at $T= m_X$. Here $x_f=m_X/T_f$ indicates the freezeout temperature, which is typically $x_f \in [15,20]$ for $3\to 2 $ freezeout.
For $m_{h_1} \gtrsim 3 m_X$, the Higgs contributions to the cross-section effectively decouples, and we have
\bea
\Omega_{\rm DM} \simeq 0.33 \left( \frac{x_f}{20}\right)^2 \left( \frac{10.75}{g_*}\right)^{3/4} \left(\frac{m_X/\alpha_X }{100~ \rm MeV} \right)^{3/2}\,.
\eea

\begin{figure}[t!]
  \begin{center}
      \includegraphics[height=0.45\textwidth]{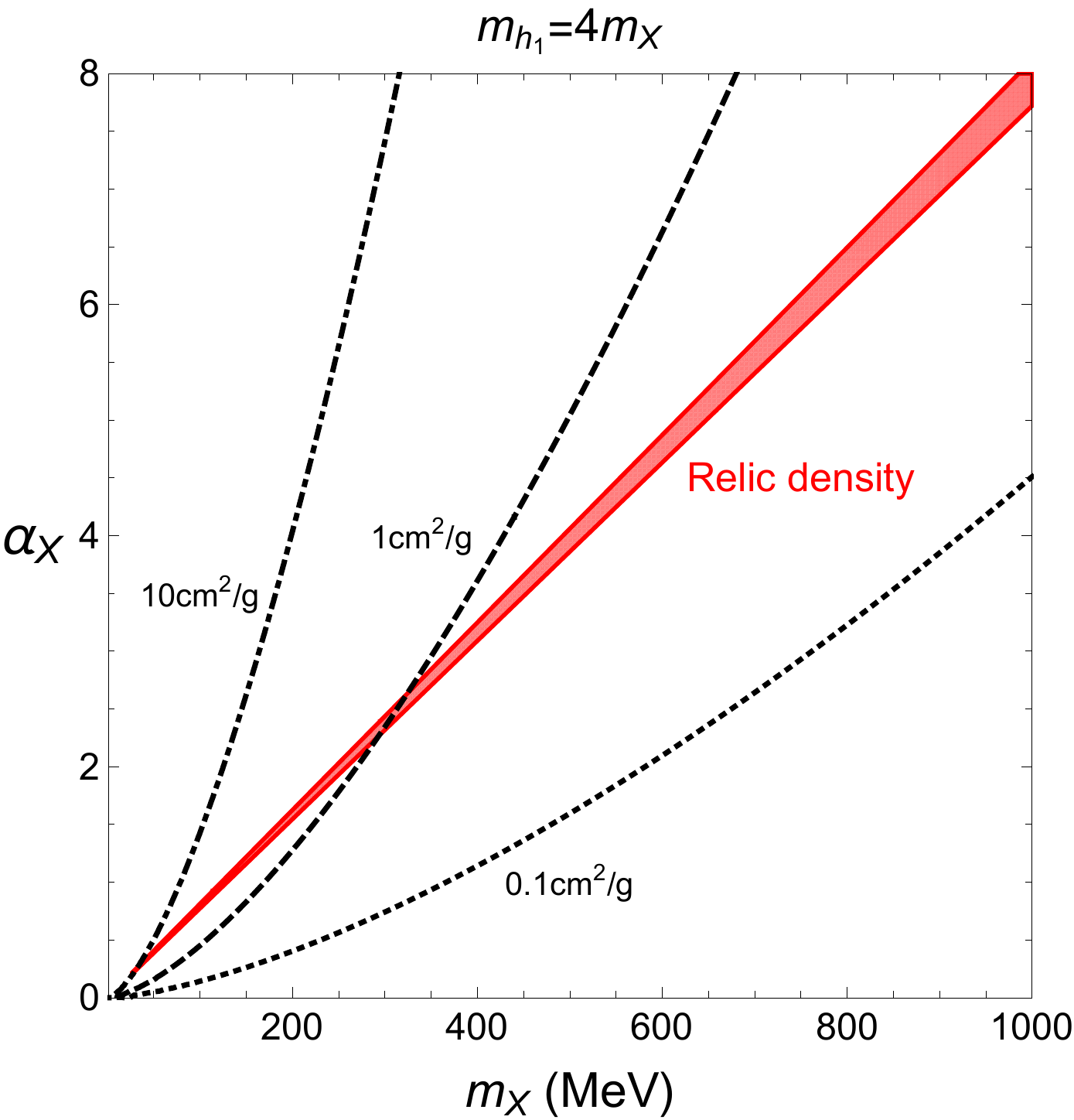}\\
             \includegraphics[height=0.45\textwidth]{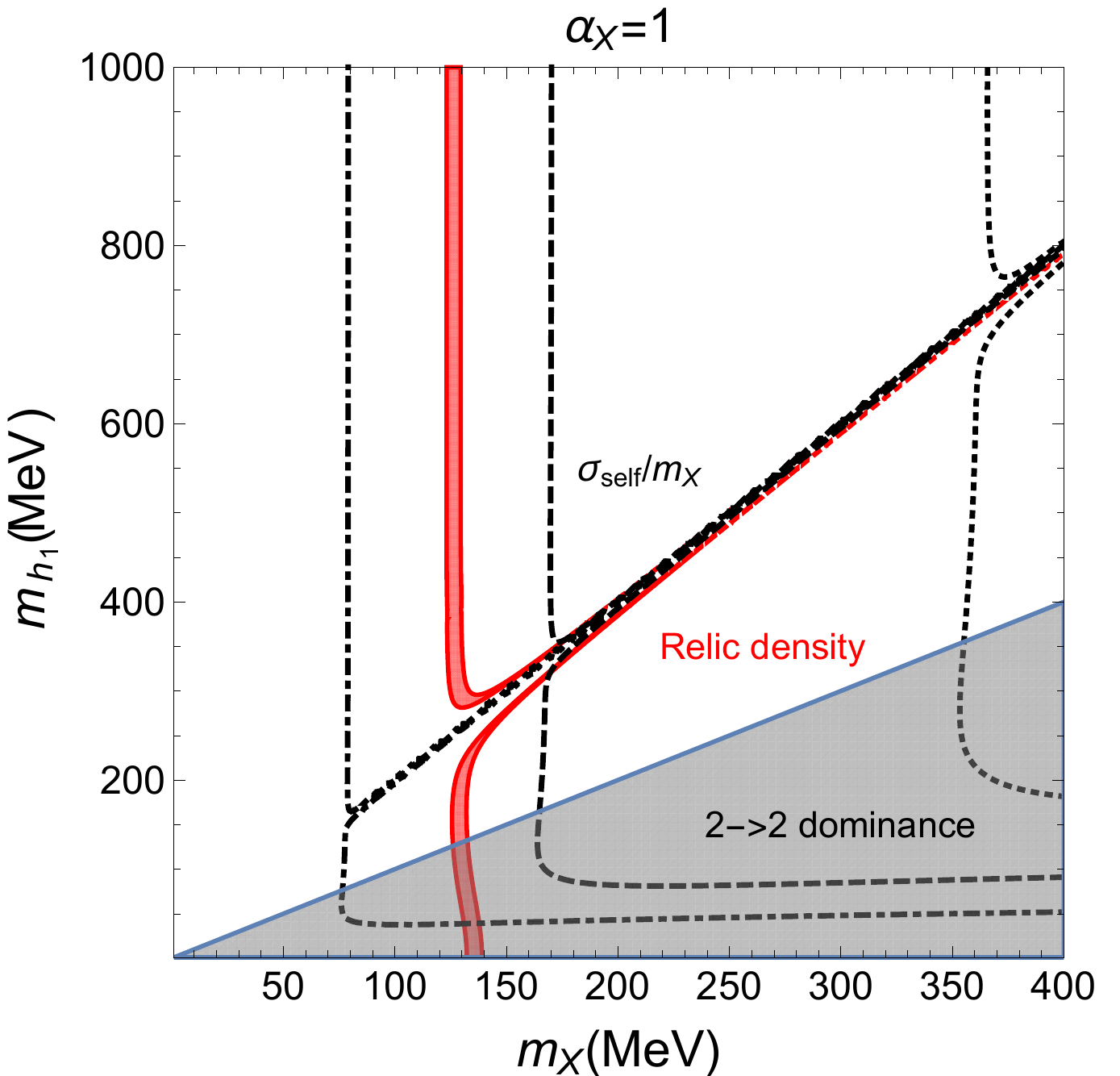}
      \includegraphics[height=0.45\textwidth]{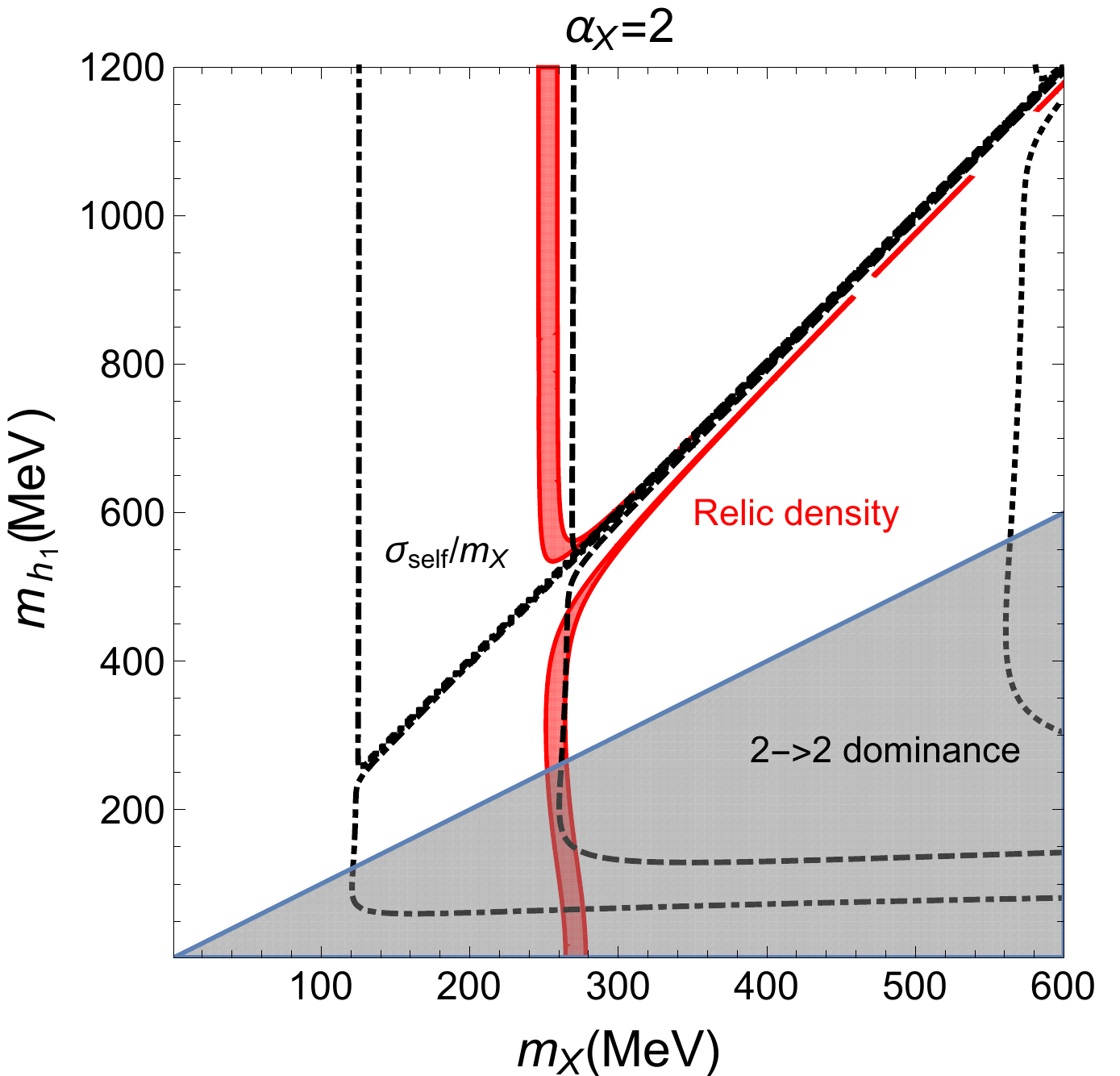}
   \end{center}
  \caption{The parameter space of vector SIMP dark matter in the $m_X$ vs. $\alpha_X\equiv g^2_X/(4\pi)$ (top) or $m_{h_1}$ (bottom), when considering $3\to2$ annihilation channels only. The Planck $3\sigma$ measurement of the relic density is show in red in all panels. Contours of the self-scattering cross section of $\sigma_{\rm self}/m_X=0.1,1,10\,{\rm cm^2/g}$ are shown in the dotted, dashed and dot-dashed curves, respectively.  We have chosen $m_{h_1}=4m_X$ on top and $\alpha_X=1, 2$ on bottom. The shaded gray regions in the lower panels are where other $2\rightarrow 2$ channels dominate over $3\to2$ processes. }
  \label{relic}
\end{figure}

In Fig.~\ref{relic} we depict the parameter space in which the measured dark matter relic density is obtained within $3\sigma$ (red region) for $\alpha_X\equiv g^2_X/(4\pi)$ ($m_{h_1}$) and $m_X$ in the upper (lower) panel. For illustration, the top panel shows the results for dark Higgs mass of $m_{h_1}=4m_X$ where no resonance enhancement is present, while in the bottom panel we fix $\alpha_X=1, 2$ and vary $m_X$ and $m_{h_1}$.

In addition to $3\to2$ annihilations, the vector SIMP dark matter undergoes self-scattering processes, which are constrained by the bullet cluster~\cite{Clowe:2003tk,Markevitch:2003at,Randall:2007ph} and by elliptical halo shapes~\cite{Rocha:2012jg,Peter:2012jh}. Away from a resonance, the self interacting cross-section is
\bea
\sigma_{\rm self}
&=&\frac{g^4_X}{1152\pi m_{h_1}^4m_X^2(m_{h_1}^2-4m_X^2)^2}
\Big(520m_{h_1}^8-4208m_{h_1}^6m_X^2+8801m_{h_1}^4m_X^4 \nonumber \\
&&-1200m_{h_1}^2m_X^6+320m_X^8\Big)\,.
\eea
A simple approximation can be derived in the limit $m_{h_1} \gg m_X$ : \begin{equation}
\dfrac{\sigma_\text{self}}{m_X}\simeq \frac{65 \pi  \alpha _X^2}{9 m_X^3} \simeq 5 \alpha_X^2 \Big( \dfrac{m_X}{100\text{ MeV}} \Big)^{-3}  \text{ cm}^2 / \text{g} \qquad \qquad [m_{h_1} \gg m_X]
\end{equation}
Contours of the self-scattering cross section obeying $\sigma_{\rm self}/m_X=0.1,1,10\,{\rm cm^2/g}$ are shown in Fig.~\ref{relic} in dotted, dashed and dot-dashed lines, respectively.

We learn that away from a resonance region, vector SIMP $3\to2$ dark matter consistent with self-scattering constraints points to dark matter masses of $m_X\gtrsim {\cal O}(100~{\rm MeV})$ and strong couplings of $\alpha_X\gtrsim 1$. Indeed, strong coupling is a frequent common feature in SIMP dark matter models~\cite{Hochberg:2014dra,Lee:2015gsa,Choi:2015bya,Hochberg:2015vrg,Choi:2016tkj,Hambye:2009fg,Hambye:2008bq,Bernal:2015ova,Cline:2017tka}, though exceptions can arise ({\it e.g.} on resonance~\cite{Choi:2016hid,Choi:2017mkk}).  
Close to the resonance region, the relic density is sensitive to the dark Higgs mass, and the viable parameter space is broadened further to include larger DM masses at fixed dark gauge coupling, or smaller dark gauge couplings for fixed DM masses.

We comment that the strong gauge coupling leads to a question on the potential breakdown of perturbativity in relic density calculation. In our case, however, the $SU(2)_X$ gauge symmetry is completely broken by the VEV of the dark Higgs, and there are no light particles below the confinement or symmetry breaking scale ({\it i.e.}\/, vector SIMP mass). Therefore, given that there is no phase transition separating the Higgs phase and confining phase, namely, the complementarity between the Higgs and confining phases~ \cite{Fradkin:1978dv,Banks:1979fi,tHooft:1979yoe,Susskind:1979up,Raby:1979my,Georgi:2016qbt}, the Higgsed theory can be pushed into regions where perturbativity is questionable. Closer inspection of the issue of complementarity may be worthwhile, though is beyond the scope of this paper. 

As the dark Higgs mass approaches the DM mass, when $m_X< m_{h_1}\lesssim 1.5 m_X$, forbidden $2\to2$ annihilation channels contribute significantly to the relic density and must be included as well; we study this in the next subsection. (The regions in which $2\to2$ processes dominate the relic density are shown in shaded gray in Fig.~\ref{relic}.) As we will see, the self-scattering rate is reduced in this case, allowing smaller dark matter masses consistent with observational constraints.

\subsection{Forbidden channels}\label{ssec:forb}

When the dark Higgs is slightly heavier than the dark matter, forbidden $2 \rightarrow 2$ channels such as $X_iX_i\rightarrow h_1 h_1$ and $X_iX_j\rightarrow X_k h_1$---although kinematically inaccessible at zero temperature---can be important in determining the relic density at the time of freeze-out~\cite{Griest:1990kh,DAgnolo:2015ujb}.  For $m_X\lesssim m_{h_1}\lesssim2(1.5)m_X$, new $3\rightarrow 2$ channels such as $XXX\rightarrow X h_1 (h_1 h_1)$ open up as well so they have been already included in Fig.~\ref{relic}.  Here we discuss the effects of the forbidden channels on the relic abundance and identify the parameter space of vector SIMP dark matter that is consistent with the observed relic density when including these effects.
(This will be particularly relevant when kinetic equilibrium between the SIMP and SM sectors is obtained via the Higgs portal, as will become evident in Section~\ref{ssec:higgsportal}.)

Assuming that the forbidden channels are dominant,  the approximate Boltzmann equation is given by
\bea
\frac{dn_{\rm DM}}{dt}+3Hn_{\rm DM}&\approx &-\frac{2}{3}\langle\sigma v\rangle_{ii\rightarrow h_1 h_1} n^2_{\rm DM}+6\langle\sigma v\rangle_{h_1 h_1\rightarrow ii} (n^{\rm eq}_{h_1})^2  \nonumber \\
&&- \frac{1}{3}\langle\sigma v\rangle_{ij\rightarrow k h_1} n^2_{\rm DM} +\langle\sigma v\rangle_{k h_1\rightarrow ij}n^{\rm eq}_{h_1} n_{\rm DM}\,,
\eea
where we have assumed that $h_1$ maintains chemical and thermal equilibrium with the SM bath throughout freezeout. See the full Boltzmann equations in Eq.~(\ref{fullBoltz}).  
Detailed balance conditions at high temperature determine the annihilation cross sections for the forbidden channels in terms of the unforbidden channels,
\bea
\langle\sigma v\rangle_{ii\rightarrow h_1 h_1} &=& \frac{9(n^{\rm eq}_{h_1})^2}{(n^{\rm eq}_{\rm DM})^2} \,\langle\sigma v\rangle_{h_1 h_1\rightarrow ii}
= (1+\Delta_{h_1})^3 e^{-2\Delta_{h_1}x}  \,\langle\sigma v\rangle_{h_1 h_1\rightarrow ii}\,, \label{dbal1} \\
\langle\sigma v\rangle_{ij\rightarrow k h_1} &=& \frac{ 3n^{\rm eq}_{h_1}}{n_{\rm DM}^{\rm eq}}\, \langle\sigma v\rangle_{k h_1\rightarrow ij} = (1+\Delta_{h_1})^{3/2} e^{-\Delta_{h_1}x} \,  \langle\sigma v\rangle_{k h_1\rightarrow ij}\,,  \label{dbal2}
\eea
with $\Delta_{h_1}\equiv (m_{h_1}-m_\chi)/m_\chi$.   The cross section formulas for the allowed $2\rightarrow 2$ channels in the RHS above are given in Appendix~\ref{app:xsec}.

Denoting the allowed $2\to2$ cross sections in the RHS above by $ \langle\sigma v\rangle_{k h_1\rightarrow ij} =a$ and $ \langle\sigma v\rangle_{h_1 h_1\rightarrow ii}=b$, the DM abundance is found to be~\cite{DAgnolo:2015ujb}
\be
Y_{\rm DM}(\infty)\approx \frac{x_f}{\lambda}\, e^{\Delta_{h_1}x_f} \, f(\Delta_{h_1},x_f)
\ee
with 
\bea
f(\Delta_{h_1},x_f)&=& \bigg[ \frac{1}{3} a (1+\Delta_{h_1})^{3/2}\Big(1-(\Delta_{h_1} x_f) \, e^{\Delta_{h_1}x_f} \int^\infty_{\Delta_{h_1} x_f} dt\, t^{-1} e^{-t} \Big) \nonumber \\
&+& \frac{2}{3} b(1+\Delta_{h_1})^3 e^{-\Delta_{h_1}x_f}\Big(1-2(\Delta_{h_1}x_f)\,e^{2\Delta_{h_1} x_F} \int^\infty_{2\Delta_{h_1} x_f} dt\, t^{-1} e^{-t} \Big)  \bigg]^{-1}\,,
\eea
resulting in the relic density
\be
\Omega_{\rm DM}h^2 = 5.20\times 10^{-10}\,{\rm GeV}^{-2} \Big(\frac{g_*}{10.75} \Big)^{-1/2} \Big(\frac{x_f}{20} \Big)\, e^{\Delta_{h_1} x_f} f(\Delta_{h_1},x_f)\,.
\ee
In general, however, one must account simultaneously for both the $3\rightarrow 2$ processes and the $2\rightarrow 2$ forbidden channels in determining the dark matter relic abundance.

\begin{figure}
  \begin{center}
  \includegraphics[height=0.45\textwidth]{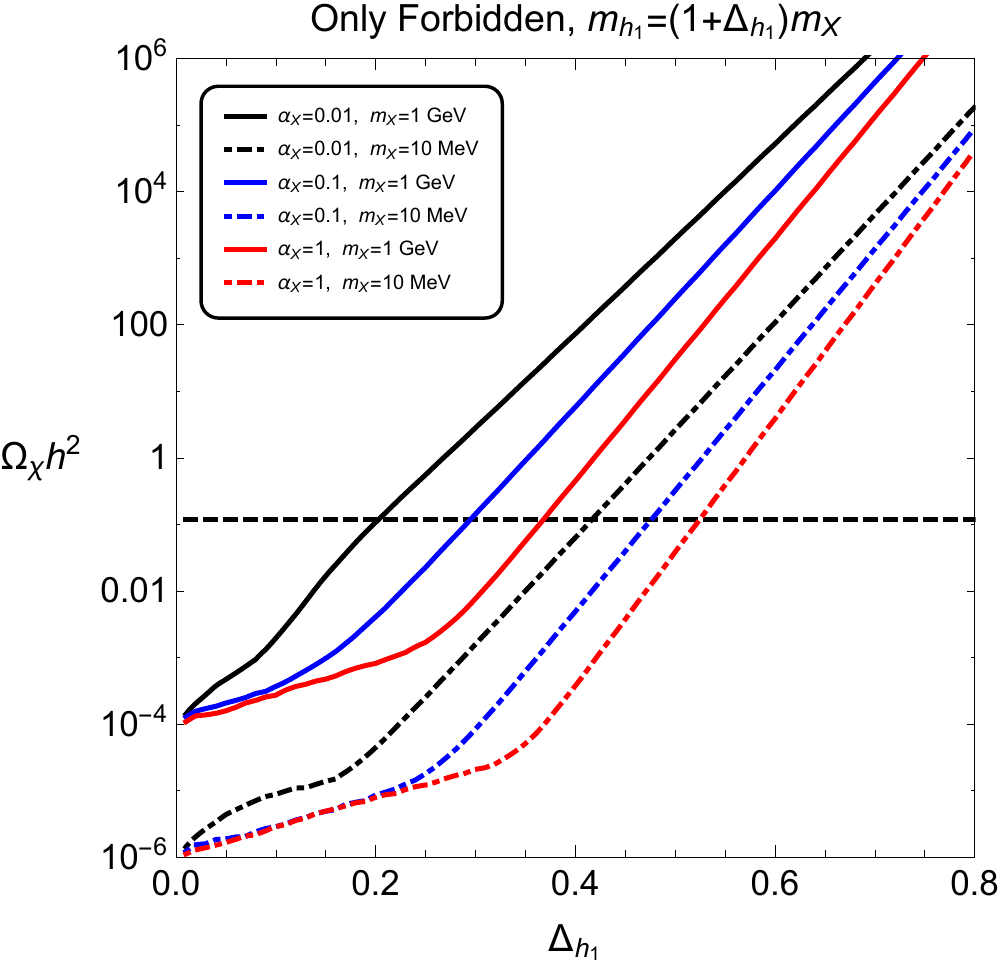}
      \includegraphics[height=0.45\textwidth]{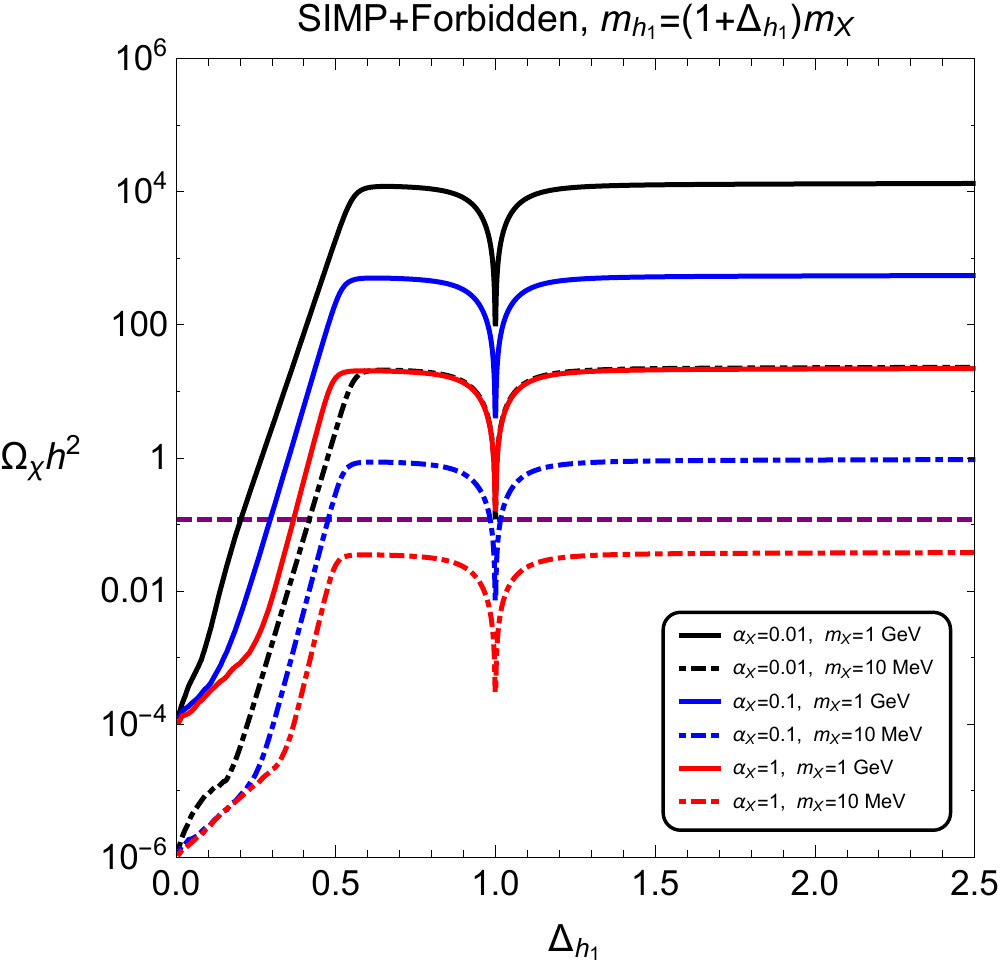}
   \end{center}
  \caption{Dark matter relic density as a function of $\Delta_{h_1}=(m_{h_1}-m_X)/m_X$, for forbidden channels only ({\bf left}) and both forbidden and SIMP channels ({\bf right}). The measured relic density is shown by the purple curve. We show the results for various illustrative values of coupling and mass: $\alpha_X=0.01,0.1,1$ and $m_X=0.1\,{\rm MeV}, 1\,{\rm GeV}$.}
  \label{fdm2}
\end{figure}

In Fig.~\ref{fdm2}, we show the dark matter relic density as a function of $\Delta_{h_1}=(m_{h_1}-m_X)/m_X$, first when including only forbidden channels ({\bf left panel}) and then when taking both forbidden and SIMP channels into account ({\bf right panel}). We have varied $\alpha_X$ and $m_X$ between $0.01-1$ and $10\;{\rm MeV}-1\,{\rm GeV}$, respectively.
We learn that forbidden channels play an important role for $\Delta_{h_1}\lesssim 0.5$, where the observed relic density can be achieved over a broad range of couplings $\alpha_X$ and masses $m_X$. As the mass difference increases, $3\to2$ SIMP annihilations begin dominating the relic abundance as a saturated value for mass differences $\Delta_{h_1}\gtrsim 0.5$. We note that the importance of the forbidden semi-annihilation channels for $\Delta_{h_1}\lesssim 0.5$, in contrast to the naive expectation from the Boltzmann suppression factors of $\Delta_{h_1}\lesssim 1$, is due to a large numerical factor in the SIMP $3\to2$ annihilation cross section.

\section{Kinetic equilibrium}\label{sec:portal}

In order of the SIMP mechanism to be viable, we require that the SIMP sector efficiently dumps entropy into the SM bath. In the proposed framework, this can be achieved either by a Higgs or $Z'$ portal between the vector SIMPs and SM particles. After a general discussion of the relevant Boltzmann equation in terms of the dark sector temperature and the requirement of equilibration in Section~\ref{ssec:boltz}, we study the Higgs portal in Section~\ref{ssec:higgsportal} and the $Z'$ portal in Section~\ref{ssec:Zprimeportal}.

\subsection{Equilibration conditions}\label{ssec:boltz}

Following Ref.~\cite{Kuflik:2015isi}, we find the decoupling temperature by comparing the rate of change in kinetic energy injected by the $3\to 2$ annihilations compared to the kinetic energy lost due to elastic scattering.  When the $3\to 2$ occurs, the mass of one dark particle is converted to the kinetic energy of the 2 outgoing particles. These  particles quickly scatter off the dark matter particles, and distribute the energy to the dark bath. Thus, the $3\to 2$  annihilations maintain chemical equilibrium in the DM gas, while releasing kinetic energy per particle
\beq
\dot{K}_{3\to 2} = m_{\rm DM}\, \frac{ \dot{n}_{\rm DM}}{n_{\rm DM}}\simeq  -m_{\rm DM}^2 H T^{-1}.\label{hubbleloss}
\eeq

Elastic scattering processes transfer this excess kinetic energy to the SM gas at a rate
\beq
\dot{K}_{\rm el} = \frac{1}{2 E_{p}} \sum_i \frac{g_i d^3 k_i}{(2\pi)^3 2 E_i} \frac{ d^3 k^\prime_i}{(2\pi)^3 2 k^\prime_i}  \frac{ d^3 p^\prime}{(2\pi)^3 2 p^\prime} \delta^4(p+k_i - p^\prime - k^\prime_i) \overline{|\mathcal{M}|^2} \left(E_{p} -E_{p^\prime} \right).
\eeq
Here, the sum is taken over the species $i$ in the relativistic plasma with initial(final) momentum $k(k^\prime)$, $p(p^\prime)$ is the dark matter initial(final) momentum. In the limit of $|{\vec k}|\ll m_{\rm DM}$, the change in kinetic energy can be written in terms of the momentum relaxation rate, $\gamma(T)$
\beq
\dot{K}_{\rm el} \simeq T \gamma(T)=\sum_i \frac{g_i}{6m_{\rm DM}} \int^\infty_0 \frac{d^3 {\vec k}}{(2\pi)^3} \, f_i (1\pm f_i) \frac{|{\vec k}|}{\sqrt{{\vec k}^2+m^2_i}}\,\int^0_{-4k^2} dt(-t) \frac{d\sigma_{Xi\rightarrow Xi}}{dt}, \label{elasticloss}
\eeq
where $t$ is the squared momentum transfer between DM and the relativistic species. The differential elastic scattering cross section is given by
\bea
 \frac{d\sigma_{Xi\rightarrow Xi}}{dt} =\frac{1}{64\pi m^2_{\rm DM}k^2}\, \overline{|{\cal M}_{Xi\rightarrow Xi}|^2}.
\eea

The decoupling occurs when the DM-to-SM energy transfer can no longer keep up with the kinetic energy production; equating Eq.~(\ref{hubbleloss}) with Eq.~(\ref{elasticloss}), we have
\beq
\gamma(T_{\rm KD}) \simeq H(T_{\rm KD}) \frac{m_{\rm DM}^2}{T_{\rm KD}^2} \label{eq:eqcond},
\eeq
where $H=0.33 g^{1/2}_* T^2/M_{\rm Pl}$ with $g_*=10.75$ the effective relativistic number of species for $1\,{\rm MeV}\lesssim T\lesssim 100\,{\rm MeV}$ and $M_{\rm Pl}=2\times 10^{18}$~GeV the Planck scale. In what follows we use Eq.~\eqref{eq:eqcond}, evaluated at $T_{\rm KD} = m_{\rm DM}/20$, to place a lower bound on the interactions between the vector SIMPs and the SM particles, needed to achieve the correct DM abundance. The ELDER DM curve~\cite{Kuflik:2015isi,Kuflik:2017iqs}, corresponds to $T_{\rm KD} \simeq m_{\rm DM}/15$, where the relic abundance is determined by the elastic scattering rate.

The dark matter can also thermalize with the SM, if the dark matter maintains equilibrium with the dark Higgs, while the dark Higgs maintains equilibrium with the SM bath via decay and inverse decays into SM fermions. The dark Higgs should be heavier than the dark $X$-bosons, or else the dark matter will efficiently annihilate into dark Higgs, effectively becoming a WIMP-like scenario. However, if the dark Higgs is much heavier than the dark matter, then the dark Higgs abundance will have been sufficiently depleted and it will not be able to maintain equilibrium between the two sectors.
This pushes the spectrum to a forbidden regime, $m_X< m_{h_1} \lesssim 1.5 m_X$, where the dark matter can annihilate into dark Higgses, but with a large Boltzmann suppression. At the time right before freezeout, both the semi-annihilation $XX \to X h_1 $ and self-annihilation $XXX \to XX$ processes will be active for large gauge coupling. The dark sector will be in thermal equilibrium with vanishing chemical potential. Thus in order for freezeout to occur one just needs to check that the dark Higgs can deplete the density in the dark sector fast enough up until freezeout,
\beq
n^{\rm eq}_{h_1}(T_{\rm FO}) \Gamma_{h_1 \to \rm SM} >  H(T_{\rm FO}) \left[n_{X }^{\rm eq}(T_{\rm FO})+ n_{h_1}^{\rm eq}(T_{\rm FO}) \right]\,. \label{eq:eqcond2}
\eeq
We use the above condition on the dark Higgs decay rate in the case that vector SIMPs are in kinetic equilibrium through the scattering with the dark Higgs.

\subsection{Higgs portal}\label{ssec:higgsportal}

The coupling $\lambda_{\Phi H}$ present in Eq.~\eqref{eq:Vhiggs} leads to mixing between the SM and dark Higgs, which enables a Higgs portal between the dark and visible sectors.

In the presence of Higgs-portal induced mixing between the SM and dark Higgs, the SM Higgs can decay invisibly into a pair of dark matter particles, with decay rate given by Eq.~\eqref{eq:h2decay}:
\bea
\Gamma(h_2\rightarrow X X)= \frac{3\sin^2\theta m^3_{h_2}}{32 \pi v^2_X}\,\bigg(1-\frac{4m^2_X}{m^2_{h_2}}+\frac{12 m^4_X}{m^4_{h_2}} \bigg) \sqrt{1-\frac{4m^2_X}{m^2_{h_2}}}.
\eea
The combined VBF, $ZH$ and gluon fusion production of Higgs bosons at CMS leads to ${\rm BR}(h_2\rightarrow XX)<0.24$ at $95\%$ CL~\cite{Khachatryan:2016whc}, while the ATLAS bounds from the VBF~\cite{Aad:2015txa} and $ZH$~\cite{Aad:2014iia} modes give ${\rm BR}(h_2\rightarrow XX)<0.29$ and ${\rm BR}(h_2\rightarrow XX)<0.75$, respectively. These decays provide a strong constraint on the mixing: $\sin\theta \lesssim 10^{-5}$ for $\alpha_X\sim \mathcal{O}(1)$.

The mixing also induces direct couplings of the darks Higgs to the SM electron and muons, which in turn induces tree-level scattering of the SM of the leptons. However, the smallness of the electron Yukawa coupling and the Boltzmann-suppression of the muons at the time of freezeout combined with constraints on the Higgs invisible decay result in the elastic scattering being inefficient for thermalization.

\begin{figure}
  \begin{center}
      \includegraphics[height=0.46\textwidth]{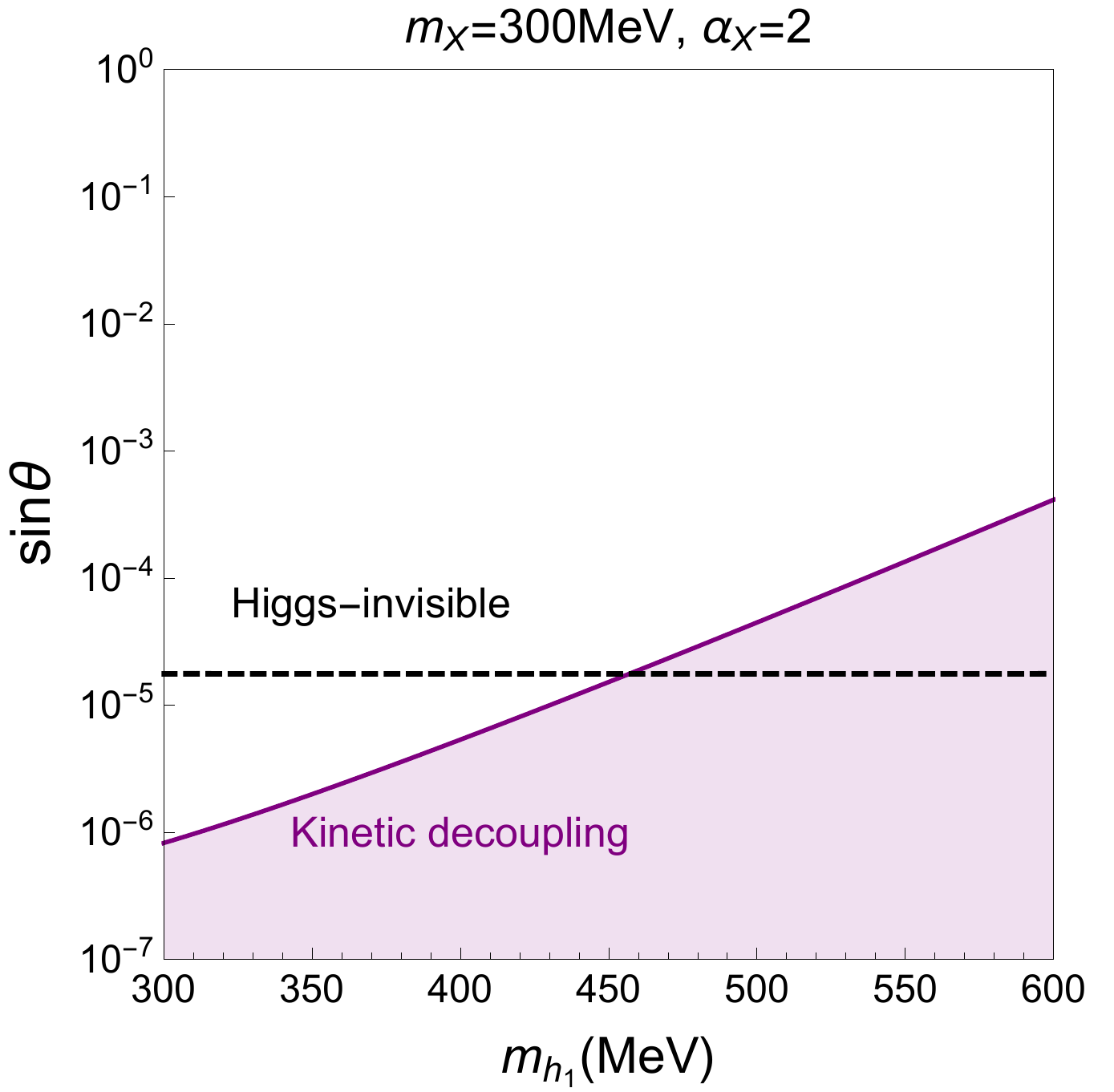}
       \includegraphics[height=0.44\textwidth]{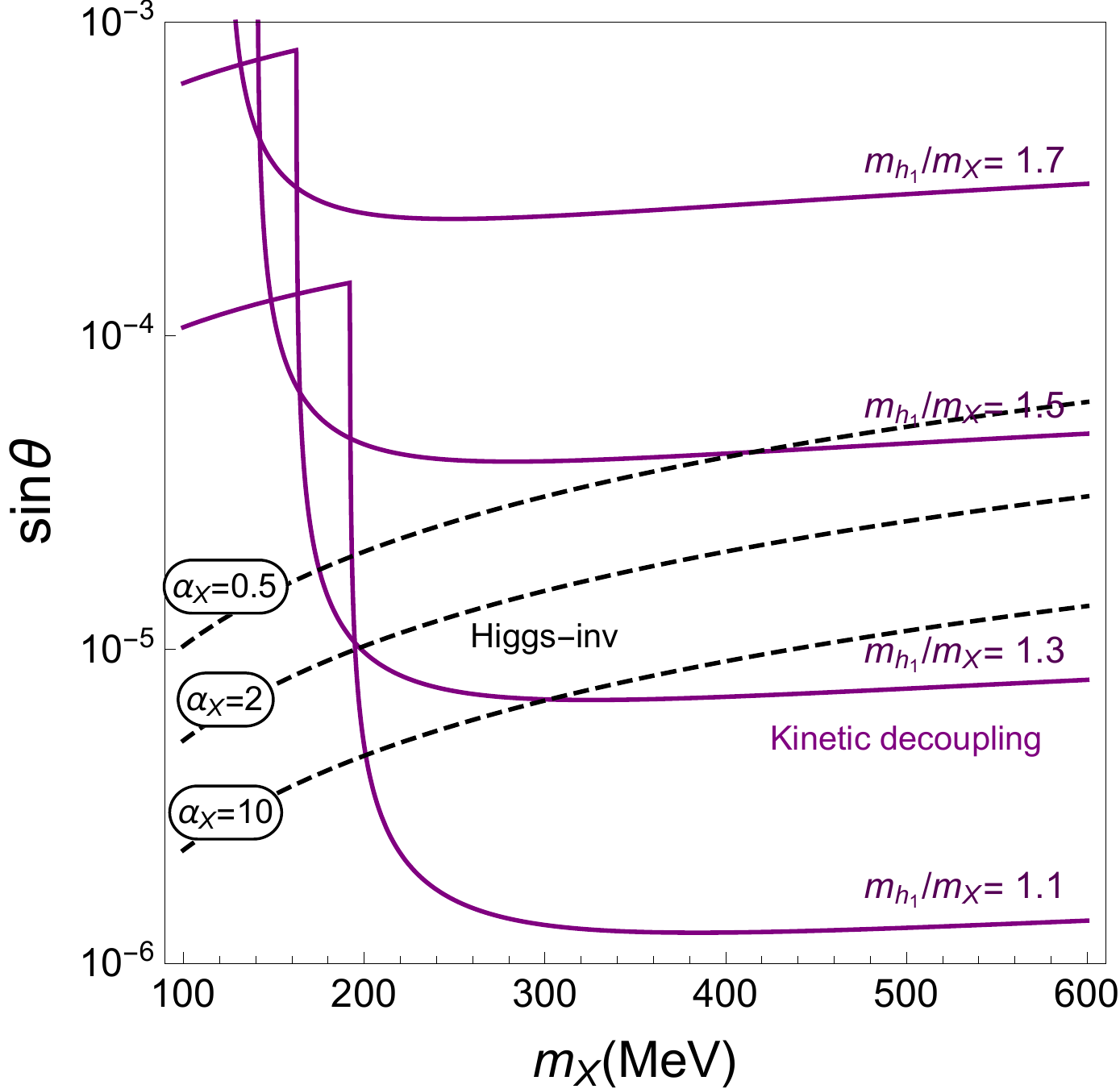}
   \end{center}
  \caption{Vector SIMPs through the Higgs portal, with DM-dark Higgs scattering and dark Higgs-SM decays. {\bf Left:} Parameter space of $m_{h_1}$ vs. $\sin\theta$ for DM-dark Higgs scattering. The shaded purple regions indicate where kinetic equilibrium between the DM and dark Higgs fails.  {\bf Right:} Parameter space of $m_X$ vs. $\sin\theta$. The purple lines are the lower bounds on $\sin\theta$ from kinetic equilibrium for fixed ratios $m_{h_1}/m_X$.  {\bf In both panels:} the dashed black curves are the upper bounds on $\sin\theta$ from Higgs invisible decays.
  }
  \label{darkH}
\end{figure}

Alternatively, if the dark Higgs is fairly light, scattering between the vector dark matter and the dark Higgs can equilibrate the dark sector, with decays and inverse decays of the dark Higgs into SM particles completing the equilibration requirement between the SIMP and SM sectors.
The momentum relaxation rate from the elastic scattering of dark matter off of dark Higgs, $X_i h_1\rightarrow X_i h_1$, is given by
\bea\label{eq:eqHDM}
\gamma(T)_{h_1}= \frac{g_{h_1} g^4_X m^2_{h_1}}{12\pi^3 m_X (m_X+m_{h_1})^2}
\bigg(\frac{m^2_{h_1}-6m^2_X}{m^2_{h_1}- 4m^2_X} \bigg)^2 T^2\,e^{-m_{h_1}/T}
\eea
where $g_{h_1}=1$.  We note that the above result is valid for $m_{h_1}(m_{h_1}-2m_X)\gtrsim p^2_{\rm DM}\sim m^2_{\rm DM} v^2_{\rm DM}$.
Plugging this into the kinetic equilibrium condition Eq.~\eqref{eq:eqcond}, we find that equilibrium between the dark Higgs and the DM is effective in most of parameter space satisfying the dark matter relic abundance.

Simultaneously, kinetic equilibrium between the dark Higgs and the SM is maintained by decays and the inverse decays of the Higgs into a pair of SM fermions,
\bea\label{eq:eqHSM}
\Gamma(h_1\to f \bar f)=\frac{m^2_f m_{h_1} \sin^2\theta}{8\pi v^2} \left(1-\frac{4m^2_f}{m^2_{h_1}}\right)^{3/2}\,.
\eea
In Fig.~\ref{darkH}, we illustrate this second requirement of equilibration between the dark Higgs and the SM, as a function of $\sin\theta$ and $m_{h_1}$ ({\bf left}) or $m_X$ ({\bf right}) for fixed $m_X$ and $\alpha_X$ ({\bf left}) or fixed ratio $m_{h_1}/m_X$ ({\bf right}). The upper bound on the mixing angle from invisible Higgs decays is indicated by the dashed black curves in both panels. Here the active thermalization process comes primarily from decays into muons when kinematically accessible, and from electrons for smaller masses.

We learn that the Higgs portal is a viable mediator between vector SIMPs and the SM when the dark Higgs is close in mass to the DM. In this regime, $2\rightarrow 2$ forbidden (semi)-annihilations channels of DM and the dark Higgs, $X_i X_j\rightarrow X_k h_1 (h_1 h_1)$, can be active and are then important contributors in determining the dark matter relic density, as discussed in Section~\ref{ssec:forb}. In this case, the semi-annihilations are also active thermalization processes within the dark sector.

We note that current limits on Higgs mixing from rare kaon- and $B$-meson decays are weaker than the bound we impose from from the Higgs invisible decay. However future beam dump or fixed target experiments, such as SHiP at CERN SPS, have the potential to probe the Higgs mixing angle further down~\cite{Alekhin:2015byh}. The allowed parameter space for the Higgs portal to vector SIMPs could then be further probed as the invisible Higgs decay constraint improves.

Before ending this subsection, we remark that a Higgs portal coupling could allow in principle for the elastic scattering of relic vector SIMP dark matter with electrons in direct-detection experiments~\cite{Essig:2011nj,Graham:2012su,Lee:2015qva,Essig:2015cda,Hochberg:2015pha,Hochberg:2015fth,Hochberg:2016ntt,Essig:2016crl,
Derenzo:2016fse,Tiffenberg:2017aac,Essig:2017kqs}.  For $m_e, m_X,m_{Z'}\gg p_{\rm DM}\simeq m_X v_{\rm DM}$, the DM-electron direct detection scattering cross section via the Higgs portal is given by
\bea
\sigma_{\rm DD}&=&\frac{\alpha_X\sin^2\theta\cos^2\theta m^4_e m^2_X}{v^2 (m_e+m_X)^2}\left(\frac{1}{m^2_{h_1}}-\frac{1}{m^2_{h_2}}\right)^2\nonumber  \\
&\approx & 4\times 10^{-50}\,{\rm cm}^2 \left(\frac{\alpha_X}{2}\right)\left(\frac{\sin\theta}{10^{-4}}\right)^2\left(\frac{1.2}{m_{h_1}/m_X}\right)^4 \left(\frac{100\,{\rm MeV}}{m_X}\right)^4\,.
\eea
The small electron Yukawa coupling suppresses the cross section substantially, yielding a currently unconstrained spin-independent direct detection cross section.

\subsection{$Z'$ portal}\label{ssec:Zprimeportal}

\begin{figure}
  \begin{center}
   \includegraphics[height=0.42\textwidth]{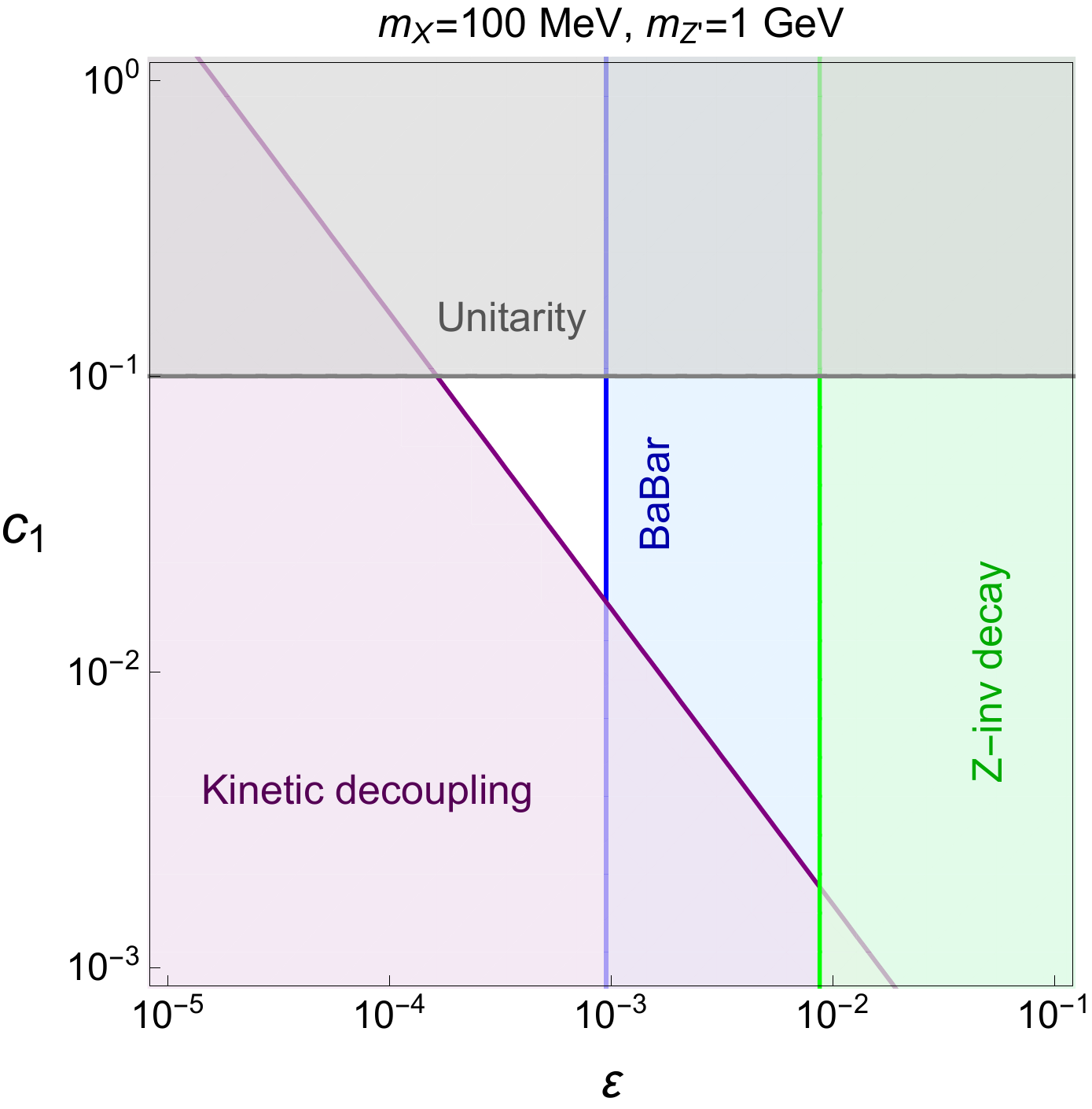}
       \includegraphics[height=0.42\textwidth]{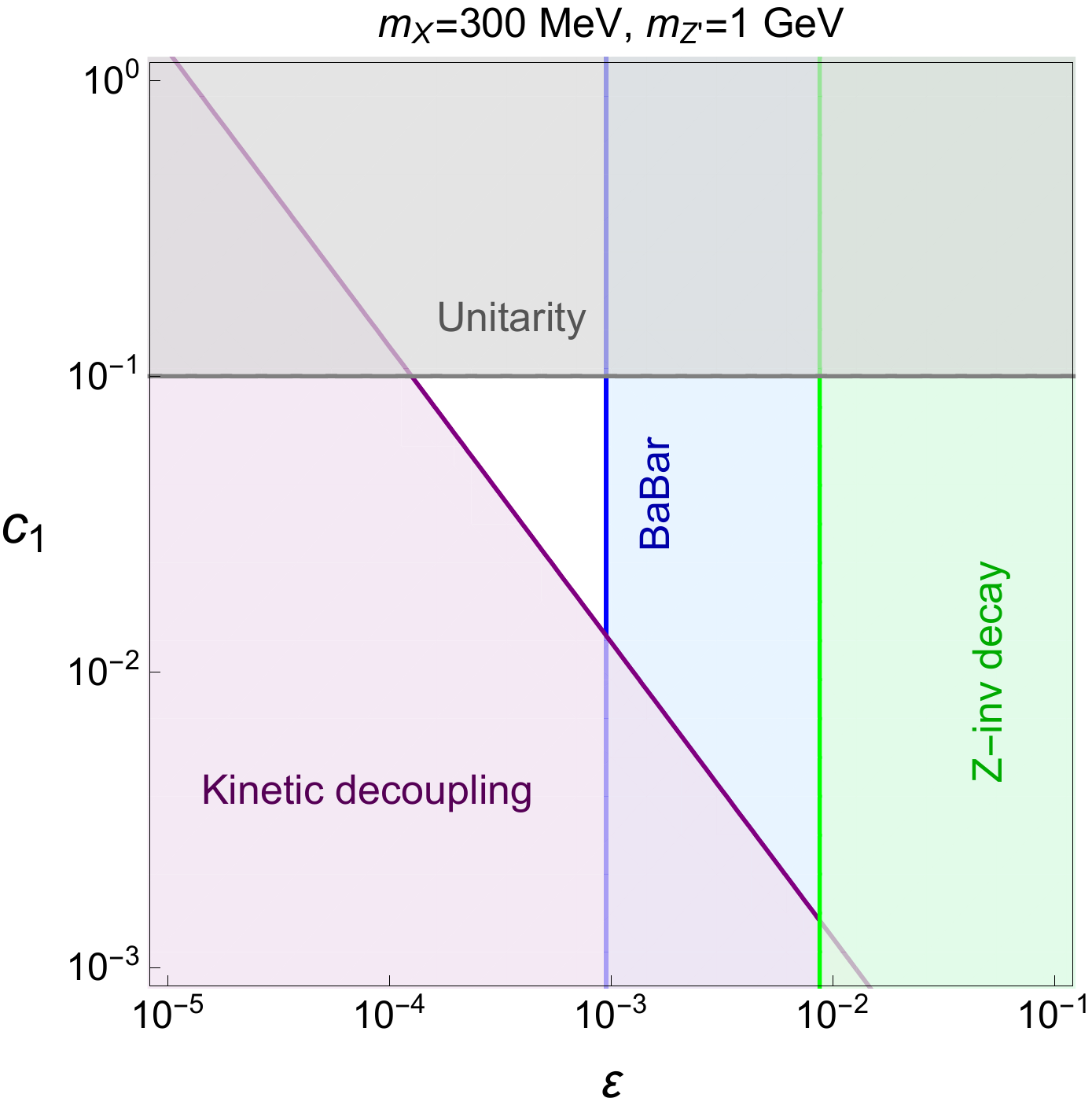}
   \end{center}
   \caption{Allowed parameter space for vector SIMPs with a $Z'$ portal in the $\varepsilon$ and $ c_1$ plane, for fixed values of DM and $Z'$ masses. We show the bounds from unitarity (brown), kinetic equilibrium (purple), the invisible width of the $Z$ boson (green)~\cite{ALEPH:2005ab} and BaBar monophoton+MET (blue)~\cite{Lees:2017lec}. Here, we took $N_f \alpha_{Z'}=1$ in Eq.~(\ref{Z-inv}) for $Z$-boson invisible decay bounds.
   }
  \label{c1-ep}
\end{figure}

 \begin{figure}
  \begin{center}
   \includegraphics[height=0.42\textwidth]{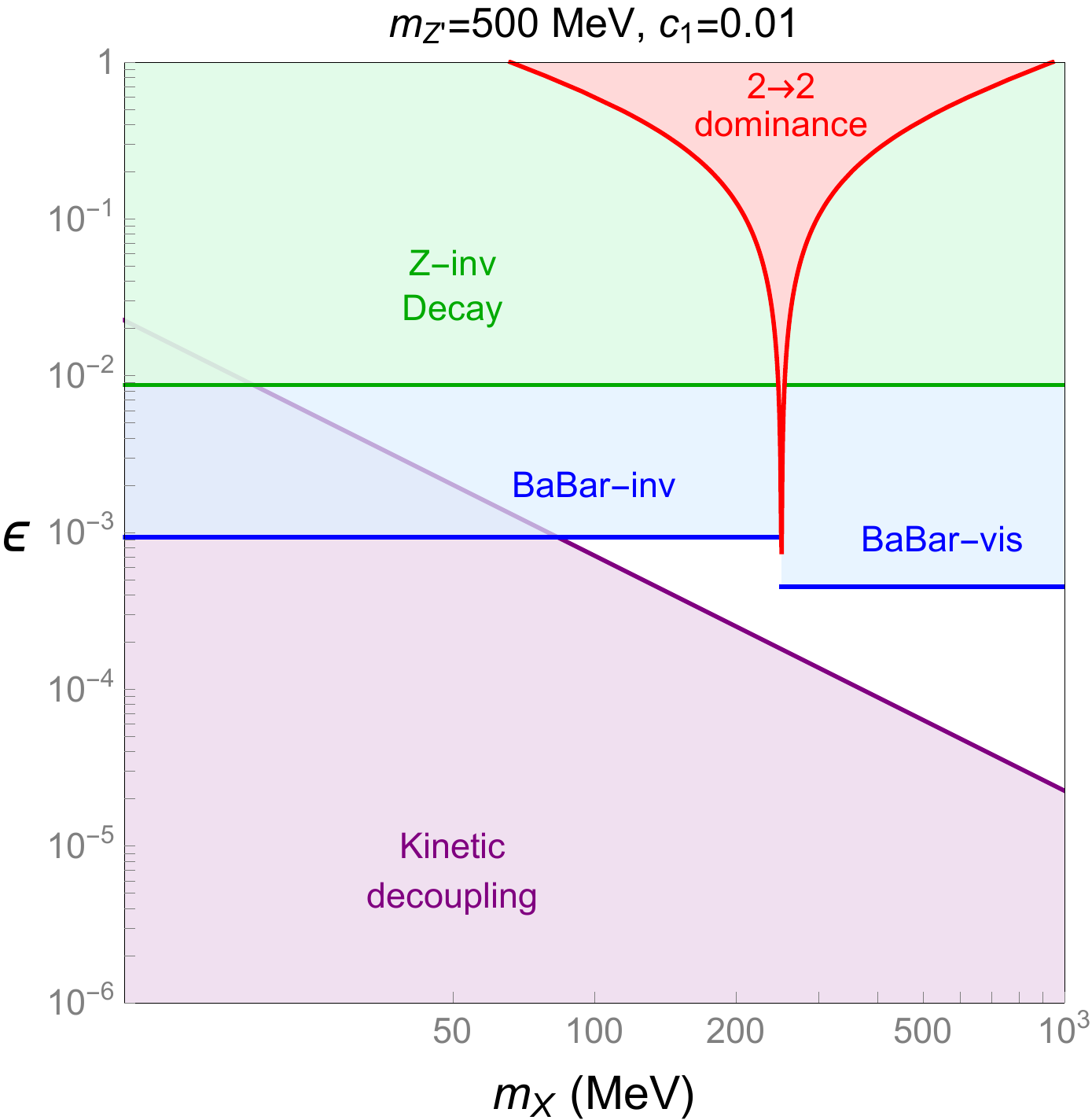}
      \includegraphics[height=0.42\textwidth]{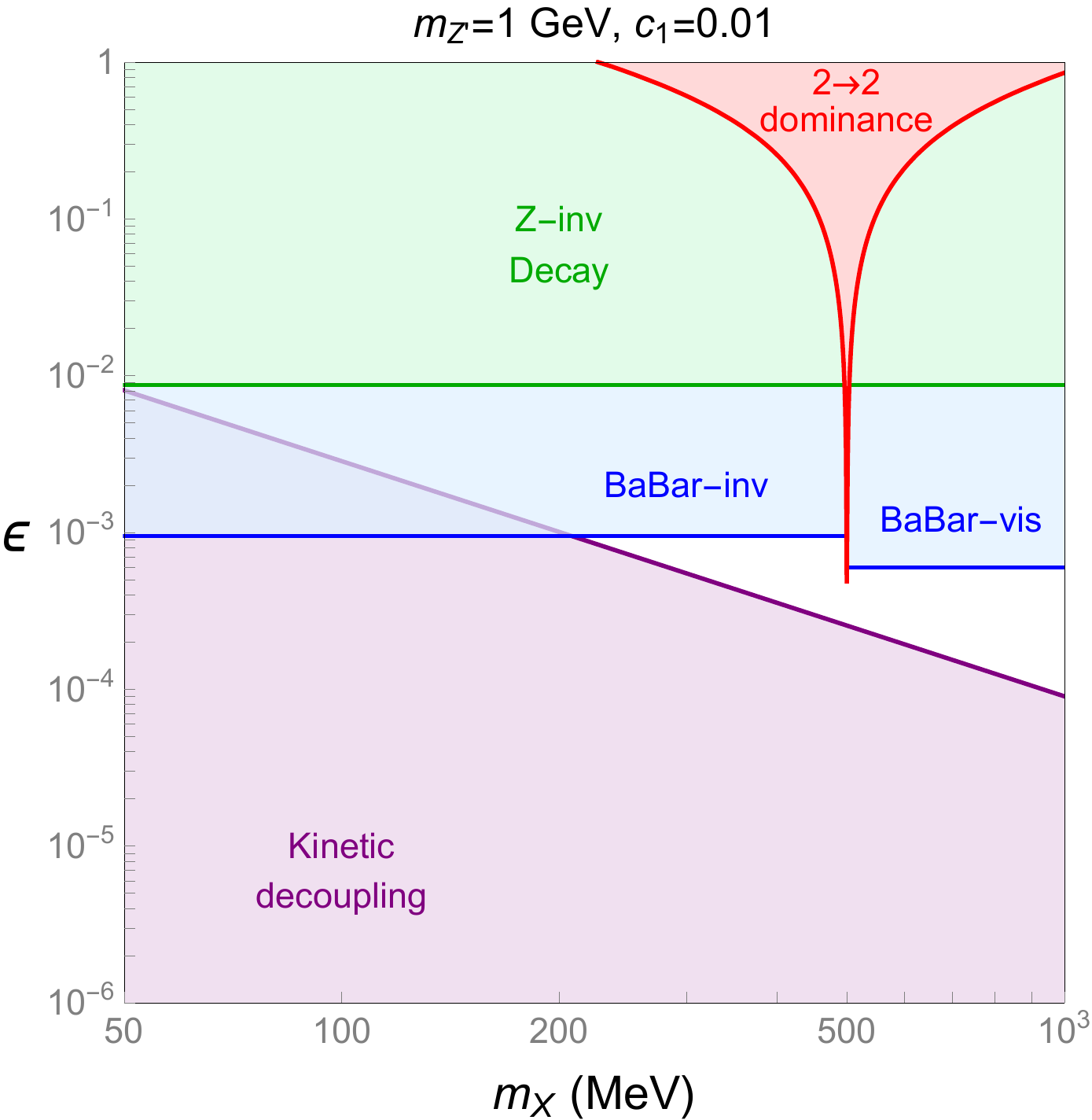} \vspace {0.5cm} \\
      \includegraphics[height=0.42\textwidth]{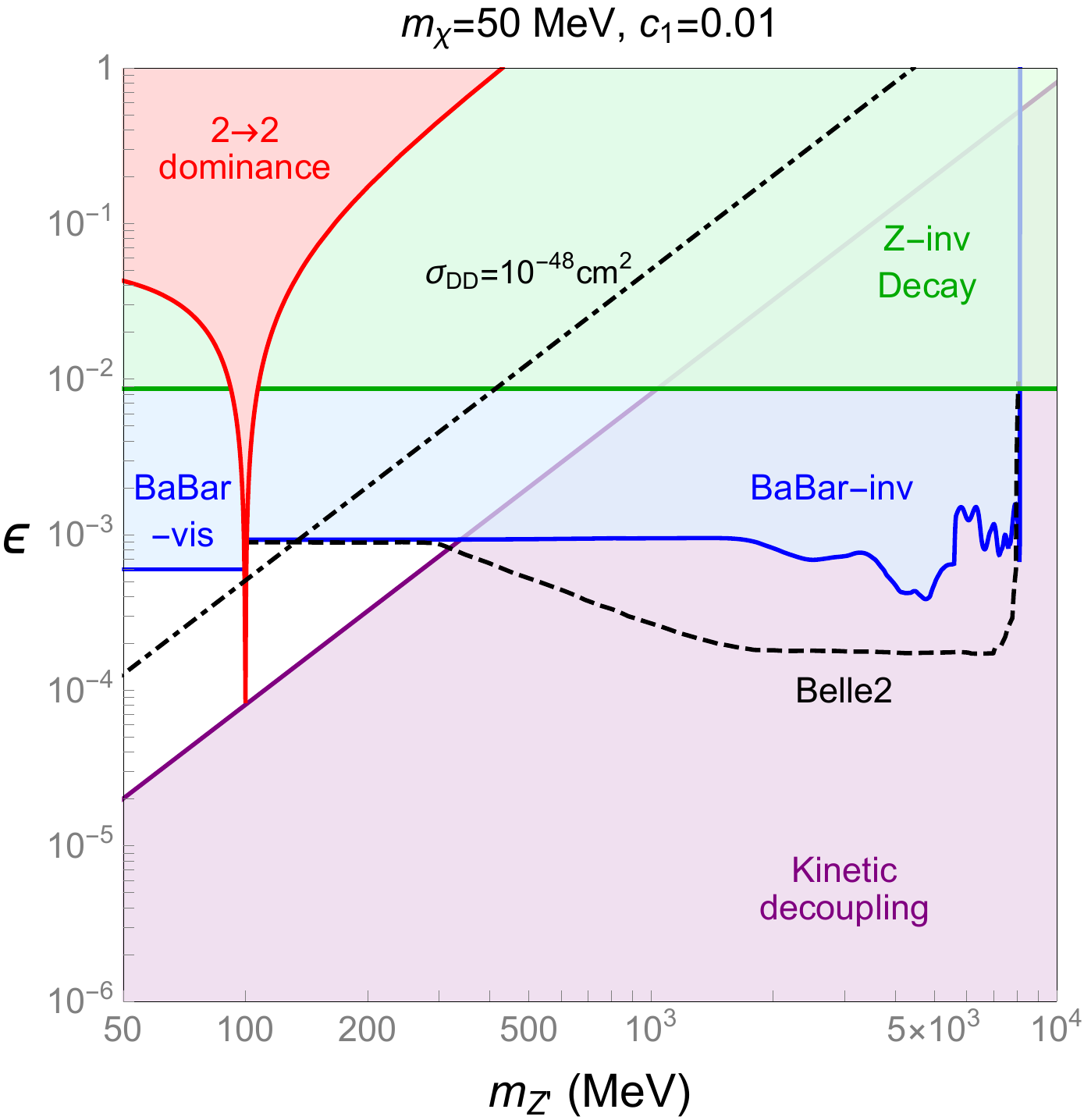}
         \includegraphics[height=0.42\textwidth]{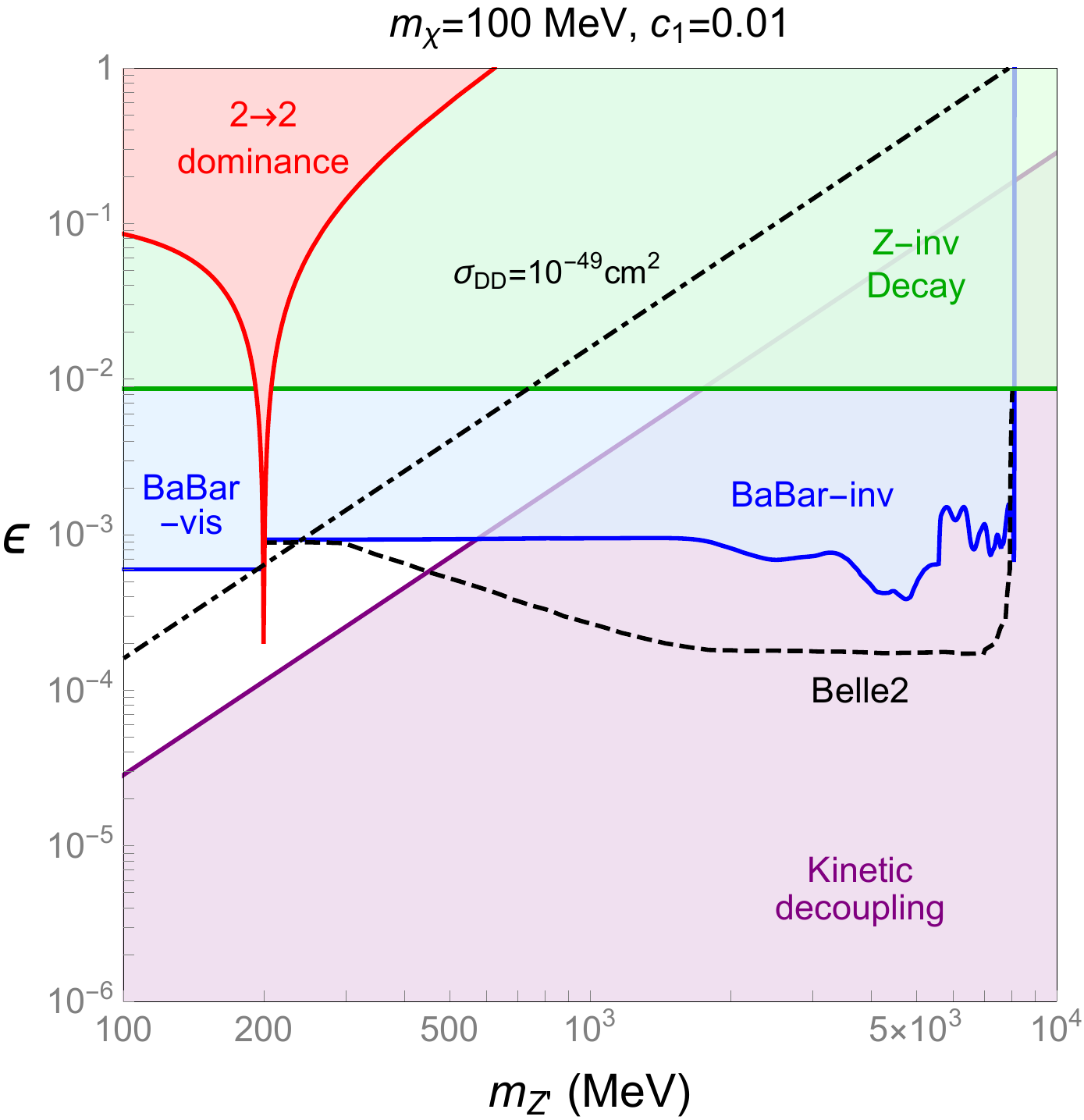}
   \end{center}
  \caption{Allowed parameter space for vector SIMPs with a $Z'$ portal. {\bf Top:} Parameter space of $m_X$ vs. $\varepsilon$ for $c_1=0.01$, for $m_{Z'}=500\,{\rm MeV}$ ({\it left}) and $1\,{\rm GeV}$ ({\it right}).
  {\bf Bottom:} Parameter space of $m_{Z'}$ vs. $\varepsilon$ for $c_1=0.01$, for $m_X=50\,{\rm MeV}$ ({\it left}) and $100\,{\rm MeV}$ ({\it right}). Here, we took $N_f \alpha_{Z'}=1$ in Eq.~(\ref{Z-inv}) for the $Z$-boson invisible decay bound.
  {\bf In all panels:} the purple region indicates where the kinetic equilibrium condition fails; the green region is excluded by the $Z$-boson invisible decay~\cite{ALEPH:2005ab}; and the red region is where the $2\rightarrow 2$ annihilation becomes dominant. BaBar searches for monophotons with MET~\cite{Lees:2017lec} and with dileptons~\cite{Lees:2014xha} exclude the blue region. The projected Belle-II reach for monophoton+MET~\cite{Essig:2013vha} is depicted in dashed blue curve.  Contours for DM-electron scattering cross section with $\sigma_{\rm DD}=10^{-48\ (49)}{\rm cm}^2$ are also shown in dot-dashed lines on the left (right) panels.  }
  \label{Zportal}
\end{figure}

Next, we explore the kinetically mixed $Z'$ portal for mediation between the SIMP and visible sectors.
We use the CS terms of Eq.~\eqref{CS1}  to couple the vector DM to the $Z'$, together with kinetic mixing between the $Z'$ and the SM hypercharge. The momentum relaxation rate for vector DM scattering with electrons via the $Z'$ portal is given by
\bea
\gamma(T)_{Z^\prime} =\frac{1240\pi^3   c_1^2e^2\varepsilon^2}{567 m_X m_{Z'}^4}\,T^6\,, \label{gTeven}
\eea
and one imposes Eq.~\eqref{eq:eqcond} for kinetic equilibrium.

The resulting allowed parameter space is depicted in Fig.~\ref{c1-ep} as a function of kinetic mixing $\varepsilon$, for fixed DM and $Z'$ masses. The gray region is excluded by the unitarity bound on the CS term, and the kinetic equilibrium condition fails in the purple region. The LEP bound on the invisible decay width of the $Z$-boson, $\Gamma_{\rm inv}<3\,{\rm MeV}$~\cite{ALEPH:2005ab} is shown in green, where we have assumed the dominant mode is into dark fermions that generate the CS coupling [as would be the case in generic UV completions with $N_f \alpha_{Z'}=1$ in Eq.~(\ref{Z-inv})] in both plots. The BaBar constraint from invisible decays~\cite{Lees:2017lec} is shown in blue (with a similar-sized constraint from the beam dump experiment NA64 at CERN SPS~\cite{Banerjee:2016tad}).

In Fig.~\ref{Zportal} we further show the allowed parameter space in $\varepsilon$ and $m_X$ (top panels) or $m_{Z'}$ (lower panels), for fixed values of the CS coefficient and of $m_{Z'}$ or $m_X$, respectively. Here, kinetic equilibrium is not maintained in the purple region; $2\to2$ processes are dominant over $3\to2$ processes in the red region; invisible $Z$-decay limits~\cite{ALEPH:2005ab} are imposed in green [where we have assumed the dominant mode is into dark fermions with $N_f \alpha_{Z'}=1$ in Eq.~(\ref{Z-inv})]; and constraints from BaBar invisible~\cite{Lees:2017lec} and visible~\cite{Lees:2014xha} searches are shown in blue. The projected reach of Belle-II into the parameter space is shown in the dashed blue curve~\cite{Essig:2013vha}.
As is evident, vector SIMPs through the $Z'$ portal can be achieved in an experimentally viable parameter space.

Concerning direct-detection, we note that the $Z'$ portal coupling of vector SIMPs via the CS term gives rise to a $p$-wave velocity-suppressed elastic cross section off electrons. As a result, the spin-independent cross section between vector SIMPs and electrons via the $Z'$ portal is highly suppressed, in contrast to the case of scalar SIMPs~\cite{Choi:2015bya}. For $m_e, m_X,m_{Z'}\gg p_{\rm DM}\simeq m_X v_{\rm DM}$, the DM-electron scattering cross section with $Z'$ portal is given by
\bea
\sigma_{\rm DD}&=&\frac{16 c_1^2 \varepsilon^2 \alpha_{\rm em} m^2_e }{3 m^4_{Z'}} \frac{(m^2_X+2m_e m_X-m^2_e) m_X^2}{(m_X+m_e)^4}\, v^2_{\rm DM} \nonumber \\
&\approx & 6\times 10^{-51}\,{\rm cm}^2 \left(\frac{c_1}{0.01} \right)^2\left(\frac{\varepsilon}{10^{-3}}\right)^2 \left(\frac{500\,{\rm MeV}}{m_{Z'}}\right)^4\left(\frac{v_{\rm DM}}{10^{-3}}\right)^2\,.
\eea
For illustration, contours of DM-electron scattering with $\sigma_{\rm DD}=10^{-48\ (49)}{\rm cm}^2$ are depicted in the lower left (right) panel of Fig.~\ref{Zportal}.

We learn that a kinetically mixed $Z'$ with CS couplings can successfully mediate interactions between the SIMP and SM sectors, consistent with all experimental constraints. We expect that future experiments such as Belle-II~\cite{Essig:2013vha} and potentially measurements at LHCb~\cite{Ilten:2016tkc}
can further probe the allowed parameter space for vector SIMPs with a vector portal.

\section{Conclusions}\label{sec:conc}

We have considered a spontaneously broken $SU(2)_X$ gauge theory in the hidden sector as an economical realization of vector SIMP dark matter. Kinetic equilibrium between the dark and visible sectors can be obtained via a Higgs portal in a minimal model or through a $Z'$-portal in an extended model with an additional $U(1)_{Z'}$ and its non-abelian Chern-Simons term. We have identified the parameter space for the $SU(2)_X$ gauge coupling and dark matter mass by taking into account the observed relic density as well as the self-scattering cross section. The kinetic equilibrium condition in combination with a variety of experimental constraints restrain the Higgs mixing or gauge kinetic mixing to a region that could be probed in current and planned experiments at the intensity frontiers.

\section*{Acknowledgments}

The work of HML and SMC is supported in part by Basic Science Research Program through the National Research Foundation of Korea (NRF) funded by the Ministry of Education, Science and Technology (NRF-2016R1A2B4008759). The work of SMC is supported in part by TJ Park Science Fellowship of POSCO TJ Park Foundation. The work of YH is supported by the U.S. National Science Foundation, grant NSF-PHY-1419008, the LHC Theory Initiative, and by the Azrieli
Foundation. EK is supported in part by the NSF under Grant No. PHY-1316222, and by a Hans Bethe Postdoctoral Fellowship at Cornell. EK would like to thank the Korea Institute for Advanced Study for hospitality, where the collaboration was initiated. YH and EK thank the Mainz Institute for Theoretical Physics for hospitality during the completion of this work.
MP and YM want to thank the Spanish MICINN's Consolider-Ingenio 2010 Programme under grant Multi-Dark {CSD2009-00064}, the contract  { FPA2010-17747}, the France-US PICS no. 06482 {,  partial support from  the European Union's Horizon 2020 research and innovation programme (under the Marie Sklodowska-Curie grant agreements No 690575 and No 674896)}, and  the ERC advanced grants.
HM was supported by the U.S. DOE under
Contract DE-AC02-05CH11231, and by the NSF under
grants PHY-1316783 and PHY-1638509. HM was also
supported by the JSPS Grant-in-Aid for Scientific Research (C) (No. 26400241 and 17K05409), MEXT Grant-in-Aid for Scientific Research on Innovative Areas (No.
15H05887, 15K21733), and by WPI, MEXT, Japan.  HM also thanks Dorota Grabowska for useful discussions.

\begin{appendix}

\section{The Chern Simons term}\label{app:CS}

\def\theequation{A.\arabic{equation}}

\setcounter{equation}{0}

In this section, we discuss the origin of the generalized CS terms in a concrete model with dark fermions for a UV completion.
Furthermore, we show that the effective CS terms and the general $Z'-X-X$ interactions can be derived from manifestly gauge invariant operators at low energy.

Suppose that there is a set of light fermions charged under $SU(2)_X\times U(1)_{Z'}$ such as
\beq
l=(2,+1), \quad {\tilde l}=(2,+1), \quad e^c=(1,-1),  \quad {\tilde e}^c=(1,-1).
\eeq
along with a heavy dark fermions with opposite $U(1)_{Z'}$ charges ($L,\tilde{L},E^c, \tilde{E}^c$) that cancel the anomalies. With dark Higgs fields of charges $\Phi=(2,0)$ and $S=(1,-2)$ , then $SU(2)$ vector-like and chiral masses from terms
\beq
S \,l \,{\tilde l}+ S^* e^c\,{\tilde e}^c +
\Phi\, l \, e^c+ {\tilde \Phi} \,{\tilde l}\, {\tilde e}^c
\eeq
where ${\tilde \Phi}=i\tau^2\Phi^*$,  are generated after $SU(2)_X\times U(1)_{Z'}$  spontaneous symmetry breaking\footnote{The $SU(2)_X$ gauge bosons masses are degenerate at tree level and receive small loop corrections due to the mass splitting between the members of each doublet fermion, that is proportional to chiral fermion mass.  If the mass splitting between $SU(2)_X$ gauge bosons is smaller than $10\%$ of DM mass, all the SIMP processes are still active and dominant and the vector dark matter remains stable for heavy fermions. One can check explicitly in the example with vector-like dark fermions that there is no $X_3-Z'$ mixing generated at loop level, so there is no issue of dark matter instability. }.

When integrating out the light fermions, the non-decoupling portion of the one-loop triangle diagrams gives an effective CS term
\beq
{\cal L}_{\rm CS,EFT} = \frac{N_f g_{Z'}\alpha_X}{4\pi} \frac{m^2_X}{m^2_f}  \epsilon^{\mu\nu\rho\sigma} Z'_\mu {\vec X}_\nu\cdot (\partial_\rho {\vec X}_\sigma - \partial_\sigma {\vec X}_\rho) ,
\eeq
where  $N_f$ being the number and mass of light fermion generations of mass, $m_f$.
For instance, for $\alpha_X=1(4)$, $N
_f=4(1)$, $g_{Z'}\sim 0.3-3$, and $m_f\sim 4m_X-10 m_X$, we find the coefficient of the operator of Eq.~\eqref{CS1}, $c_1\simeq 0.01$.
Therefore, for $m_f \gtrsim m_X$, we can avoid additional $2\rightarrow 2$ annihilations of vector dark matter into light dark fermions, such as $XX\rightarrow f {\bar f}$, and a sizable CS term required for kinetic equilibrium can be consistently realized.

Notice here that the values of $c_1 \gtrsim 10^{-2}$ required for achieving the correct relic density in this setup imply a large multiplicity of the dark fermions or sizable gauge couplings which might drive the theory toward its non perturbative regime or the unstability of the dark higgs potential vacuum for energies of the order of the GeV scale. One could invoke more elaborate mechanisms in order to solve this potential issues but those are beyond the phenomenological considerations of this work.

If one considers only the light fermions
$l=(2,+1), e^c=(1,-1)$ and their heavy partners for anomaly cancellation, then are only chiral fermion masses due to the $SU(2)_X$ breaking. In this case, the needed CS terms are not generated. Instead, a nonzero dimension-6 interaction \bea
{\cal L}_{D6}=\frac{c_3}{M^2}\, \epsilon^{\mu\mu\rho\sigma} \partial^\lambda Z'_{\mu\nu} (X_{1,\rho\sigma}X_{2,\lambda}-X_{2,\rho\sigma}X_{1,\lambda}). \label{d6}
\eea
appears, which can also be sufficient for equilibrating the two sectors

Similarly, the effective dimension-6 operator in Eq.~(\ref{d6}) can be derived from another gauge invariant dimension-8 operator,
\bea
{\cal L}_{D8}= \frac{1}{M^4} \, \epsilon^{\mu\nu\rho\sigma} (\Phi^\dagger X_{\mu\nu} D_\lambda \Phi) \partial^\lambda Z'_{\rho\sigma}.
\eea
Then, in both cases with dimension-6 and dimension-8 operators, after the $SU(2)_X$ is broken by the VEV of the scalar doublet $\Phi$, the needed $Z'XX$ interactions are generated.

The effective approach considered in this work would be valid only for processes involving energies below the lightest dark fermion mass. Our approach is then justified for the DM freeze-out process which occurs when the DM becomes non-relativistic (i.e. for processes occuring at energies $\sim m_X \ll m_f$) and the dark fermions have already decoupled for the thermal bath. However, considering the invisible decay of the $Z$ boson leads the effective approach to fail and one has to consider the complete dark fermions degrees of freedom in the computation.

\section{Thermally averaged cross sections}\label{app:xsec}

\def\theequation{B.\arabic{equation}}

\setcounter{equation}{0}

After spin average for initial states and spin sum for the final states, the $2\rightarrow 2$ self-scattering cross sections for vector dark matter  (with notations, $ij\rightarrow ij$ meaning that $X_i X_j\rightarrow X_i X_j$, etc.),  are in the non-relativistic limit
\bea
\sigma_{ii\rightarrow ii}&=&\frac{ g^4_X m_\chi^2(15m_{h_1}^4-80m_{h_1}^2m_X^2+128m_X^4)}{384\pi m^4_{h_1}(m_{h_1}^2-4m_X^2)^2}, \\
\sigma_{ij\rightarrow ij}&=&\frac{g^4_X}{192\pi m^4_{h_1} m^2_X}\,(44m_{h_1}^4-16m_{h_1}^2m_X^2+3m_X^4), \quad i\neq j,\\
\sigma_{ii\rightarrow jj}&=&\frac{g_X^4(172m_{h_1}^4-1368m_{h_1}^2m_X^2+2723m_X^4)}{384\pi m_X^2(m_{h_1}^2-4m_X^2)^2}, \quad i\neq j.
\eea

We define the thermal average for $2\rightarrow 2$ annihilation, $\phi_1\phi_2\rightarrow \phi_3\phi_4$, as follows,
\bea
\langle\sigma v\rangle=\frac{1}{n^{\rm eq}_1 n^{\rm eq}_2}\, \frac{1}{s_i s_f}\, \int d\Pi_1 d\Pi_2 d\Pi_3 d\Pi_4 f^{\rm eq}_1  f^{\rm eq}_2 (2\pi)^4 \delta^4(p)|{\cal M}_{2\rightarrow 2}|^2  \label{22ave}
\eea
where $n^{\rm eq}_{1,2}$, $f^{\rm eq}_{1,2}$ are the number densities and occupancies of particle $1,2$ in thermal equilibrium, and $s_{i,f}$ are the symmetry factors for the initial or final states, which are $s_{i,f}=1(2)$ for two identical (different) particles, and $d\Pi_i$ are the full phase space integrals for each particle, and $|{\cal M}_{2\rightarrow 2}|^2$ is the squared amplitude for $\phi_1\phi_2\rightarrow \phi_3\phi_4$.
Then, the $2\rightarrow 2$ forbidden (semi-)annihilation cross sections (with notations, $h_1 h_1\rightarrow  ii$ meaning that $ h_1 h_1 \rightarrow X_i X_i$ and $i h_1\rightarrow jk$ meaning that $X_i h_1\rightarrow X_j X_k$) are also given by
\bea
\langle \sigma v\rangle_{h_1 h_1 \rightarrow ii} &=&\frac{m_{h_1}^2}{512\pi m_X^4}\sqrt{1-\frac{m_X^2}{m_{h_1}^2}}\Big[64\lambda_\phi^2\frac{m_X^4}{m_{h_1}^4}\Big(4-4\frac{m_X^2}{m_{h_1}^2}+3\frac{m_X^4}{m_{h_1}^4}\Big) \nonumber \\
&&-16g_X^2\lambda_\phi\frac{m_X^2}{m_{h_1}^2}\Big(4+8\frac{m_X^2}{m_{h_1}^2}-15\frac{m_X^4}{m_{h_1}^4}+12\frac{m_X^6}{m_{h_1}^6}\Big)  \nonumber \\
&&+g_X^4\Big(4+20\frac{m_X^2}{m_{h_1}^2}+11\frac{m_X^4}{m_{h_1}^4}-56\frac{m_X^6}{m_{h_1}^6}+48\frac{m_X^8}{m_{h_1}^8}\Big)\Big], \\
\langle\sigma v\rangle_{i h_1\rightarrow jk} &=&\frac{g_X^4m_{h_1}^3}{384\pi m_X^5}\Big(1+3\frac{m_X}{m_{h_1}}\Big)^{3/2}\Big(1-\frac{m_X}{m_{h_1}}\Big)^{3/2}\Big(1+\frac{m_X}{m_{h_1}}\Big)^{-1}\Big(1+2\frac{m_X}{m_{h_1}}\Big)^{-2} \nonumber \\
&&\times \Big(1+4\frac{m_X}{m_{h_1}}-4\frac{m_X^2}{m_{h_1}^2}-10\frac{m_X^3}{m_{h_1}^3}+144\frac{m_X^4}{m_{h_1}^4}+396\frac{m_X^5}{m_{h_1}^5}+297\frac{m_X^6}{m_{h_1}^6}\Big)
\eea
with $i\neq j\neq k$ in the latter case.

We define the thermal average for $3\rightarrow 2$ annihilation, $\phi_1\phi_2\phi_3\rightarrow \phi_3\phi_4$, as follows,
\bea
\langle\sigma v^2\rangle=\frac{1}{n^{\rm eq}_1 n^{\rm eq}_2 n^{\rm eq}_3}\, \frac{1}{s_i s_f}\, \int d\Pi_1 d\Pi_2d\Pi_3 d\Pi_4 d\Pi_5  f^{\rm eq}_1  f^{\rm eq}_2 f^{\rm eq}_3 (2\pi)^4 \delta^4(p) |{\cal M}_{3\rightarrow 2}|^2
\eea
where $s_{i,f}$ are the symmetry factors, which are given by $s_i=n_i !$ and $s_f=n_f!$, depending on the number of identical particles in the initial and final states, $n_i$ and $n_f$, respectively, and $|{\cal M}_{3\rightarrow 2}|^2$ is the squared amplitude for $\phi_1\phi_2\phi_3\rightarrow \phi_4\phi_5$.

The $3\rightarrow 2$ annihilation cross sections including only $SU(2)_X$ gauge interactions (with notations, $123\rightarrow 11$ meaning $X_1 X_2 X_3\rightarrow X_1 X_1$, etc) are, in the non-relativistic limit,
\bea
\langle\sigma v^2\rangle_{ijk}&\equiv&\langle\sigma v^2\rangle_{123\rightarrow 11}=\langle\sigma v^2\rangle_{123\rightarrow 22}=\langle\sigma v^2\rangle_{123\rightarrow 33} \nonumber \\
&=&\frac{5\sqrt{5}g_X^6}{331776\pi m_X^5(m_{h_1}^2+m_X^2)^2}\,(347 m^4_{h_1}+586 m_{h_1}^2 m_X^2+707 m^4_X) \nonumber \\
&&+\frac{19\sqrt{5} g^6_X}{1152\pi m_X(9m^2_X-m^2_{h_1})^2}\, \langle( v^2_1+v^2_2+v_1 v_2 \cos\theta_{12})\rangle,
\eea
\bea
\langle\sigma v^2\rangle_{iij}&\equiv &\langle\sigma v^2\rangle_{112\rightarrow 13}=\langle\sigma v^2\rangle_{113\rightarrow 12}=\langle\sigma v^2\rangle_{221\rightarrow 23}=\langle\sigma v^2\rangle_{223\rightarrow 12}  \nonumber \\
&=&\langle\sigma v^2\rangle_{331\rightarrow 23}=\langle\sigma v^2\rangle_{332\rightarrow 13} \nonumber \\
&=&\frac{5\sqrt{5}g_X^6}{2654208\pi m_X^5}\Big(14377+\frac{6m_X^2(157m_{h_1}^2-763m_X^2)}{(m_{h_1}^2-4m_X^2)(m_{h_1}^2+m_X^2)} \nonumber \\
&&+\frac{3m_X^4(5281m_{h_1}^4-18558m_{h_1}^2m_X^2+32561m_X^4)}{(m_{h_1}^2-4m_X^2)^2(m_{h_1}^2+m_X^2)^2}\Big),
\eea
\bea
\langle\sigma v^2\rangle_{iii}&=&\langle\sigma v^2\rangle_{111\rightarrow 23}=\langle\sigma v^2\rangle_{222\rightarrow 13}=\langle\sigma v^2\rangle_{333\rightarrow 12} \nonumber \\
&=&\frac{25\sqrt{5}g_X^6}{2654208\pi m_X^5}\Big(8375+\frac{362m_X^2}{m_{h_1}^2-4m_X^2}+\frac{1713m_X^4}{(m_{h_1}^2-4m_X^2)^2}\Big).
\eea
Here, we have included the $p$-wave terms in $\langle\sigma v^2\rangle_{iii}$ as they have a resonance at $m_{h_1}=3m_X$. We note that $v_1, v_2$ are the speeds of two dark matter particles in the initial states and $\theta_{12}$ is the angle between the two in the center of mass frame.

On the other hand, the $3\rightarrow 2$ annihilation cross sections including the dark Higgs (with notations, $122\rightarrow 1 h_1$ meaning $X_1 X_2 X_2\rightarrow X_1 h_1$, etc) are, in the non-relativistic limit,
\bea
\langle\sigma v^2\rangle^h_{iii}&\equiv& \langle\sigma v^2\rangle_{111\rightarrow 1 h_1}=\langle\sigma v^2\rangle_{222\rightarrow 2 h_1}=\langle\sigma v^2\rangle_{333\rightarrow 3 h_1}  \nonumber \\
&=& \frac{\sqrt{5}g_X^6 m^{16}_{h_1} C_1 (1-m^2_{h_1}/(16m^2_X))^{1/2}}{17915904\pi m_X^{10}(4m_X^2-m_{h_1}^2)^{7/2}(2m_X^2+m_{h_1}^2)^2}
\eea
with 
\begin{equation}
\begin{split}
C_1\equiv &\frac{1}{m^{16}_{h_1}}\Big( 3m_{h_1}^{16}-270m_{h_1}^{14}m_X^2+9917m_{h_1}^{12}m_X^4-187056m_{h_1}^{10}m_X^6+1952400m_{h_1}^{8}m_X^8\\
&-11318848m_{h_1}^{6}m_X^{10}+35045232m_{h_1}^{4}m_X^{12}-52110336m_{h_1}^{2}m_X^{14}+30261248m_X^{16} \Big)  \label{C1}
\end{split}
\end{equation}
and 
\bea
\langle\sigma v^2\rangle^h_{ijj}&\equiv& \langle\sigma v^2\rangle_{122 \rightarrow 1 h_1}=\langle\sigma v^2\rangle_{133\rightarrow 1 h_1}=\langle\sigma v^2\rangle_{211\rightarrow 2 h_1} =\langle\sigma v^2\rangle_{233\rightarrow 2 h_1} \nonumber \\
&=& \langle\sigma v^2\rangle_{311\rightarrow 3 h_1}=\langle\sigma v^2\rangle_{322\rightarrow 3 h_1}  \nonumber \\
&=& \frac{\sqrt{5}g_X^6m^{20}_{h_1} C_2(1-m^2_{h_1}/(16m^2_X))^{1/2}}{17915904\pi m_X^{10}(4m_X^2-m_{h_1}^2)^{7/2}(2m_X^2+m_{h_1}^2)^2(7m^2_X-m^2_{h_1})^2}
\eea
with
\begin{equation}
\begin{split}
C_2\equiv&\frac{1}{m^{20}_{h_1}}\Big( 13m_{h_1}^{20}-568m_{h_1}^{18}m_X^2+33204m_{h_1}^{16}m_X^4-724140m_{h_1}^{14}m_X^6+6743931m_{h_1}^{12}m_X^8\\
&-26087280m_{h_1}^{10}m_X^{10}+48284736m_{h_1}^{8}m_X^{12}-166749984m_{h_1}^{6}m_X^{14}+806289168m_{h_1}^4m_X^{16}\\
&-2275720192m_{h_1}^2m_X^{18}+3442229248m_X^{20} \Big). \label{C2}
\end{split}
\end{equation}
We note that the factor $1/(4m^2_X-m^2_{h_1})^4$ in the above results is the squared product of the dark Higgs propagator in $s$-channel  and the dark matter propagator in $t$-channel, which are regularized at $m_{h_1}=2m_X$ by the finite width of the dark Higgs and a nonzero dark matter velocity, respectively. 
The $3\rightarrow 2$ annihilation cross sections including two dark Higgs bosons such as $XXX\rightarrow h_1 h_1$ are $p$-wave suppressed and sub-dominant, so we don't include them here.

The Boltzmann equation for the total DM density $n_{\rm DM}$ is
\bea
\frac{dn_{\rm DM}}{dt}+3Hn_{\rm DM}&=&-\frac{2}{3}\langle\sigma v\rangle_{ii}(n_{\rm DM}^2
-(n^{\rm eq}_ {\rm DM})^2) \nonumber \\
&&-\frac{1}{9}\Big(\langle\sigma v^2\rangle_{ijk}+2\langle\sigma v^2\rangle_{iij}+\langle\sigma v^2\rangle_{iii}\Big) (n_{\rm DM}^3-n_{\rm DM}^2n^{\rm eq}_{\rm DM}) \nonumber \\
&&-\frac{2}{9}\Big(\langle\sigma v^2\rangle^h_{iii}+2\langle\sigma v^2\rangle^h_{ijj}\Big) (n_{\rm DM}^3-n_{\rm DM}(n^{\rm eq}_{\rm DM})^2 ) \nonumber \\
&&-\frac{2}{3}\langle\sigma v\rangle_{ii\rightarrow h_1 h_1} n^2_{\rm DM}+6\langle\sigma v\rangle_{h_1 h_1\rightarrow ii} (n^{\rm eq}_{h_1})^2  \nonumber \\
&&-\frac{1}{3}\langle\sigma v\rangle_{ij\rightarrow k h_1} n^2_{\rm DM} +\langle\sigma v\rangle_{k h_1\rightarrow ij}n^{\rm eq}_{h_1} n_{\rm DM}.  \label{fullBoltz}
\eea
Here, we have assumed that $h_1$ maintains chemical and thermal equilibrium with the SM bath throughout freezeout. We note that $\langle\sigma v\rangle_{ii}$ in the first line is the averaged $2\rightarrow 2$ annihilation cross section into a pair of the SM fermions, i.e. $X_i X_i \rightarrow f{\bar f}$.

\end{appendix}

\bibliographystyle{jhep}
\bibliography{lit}

\end{document}